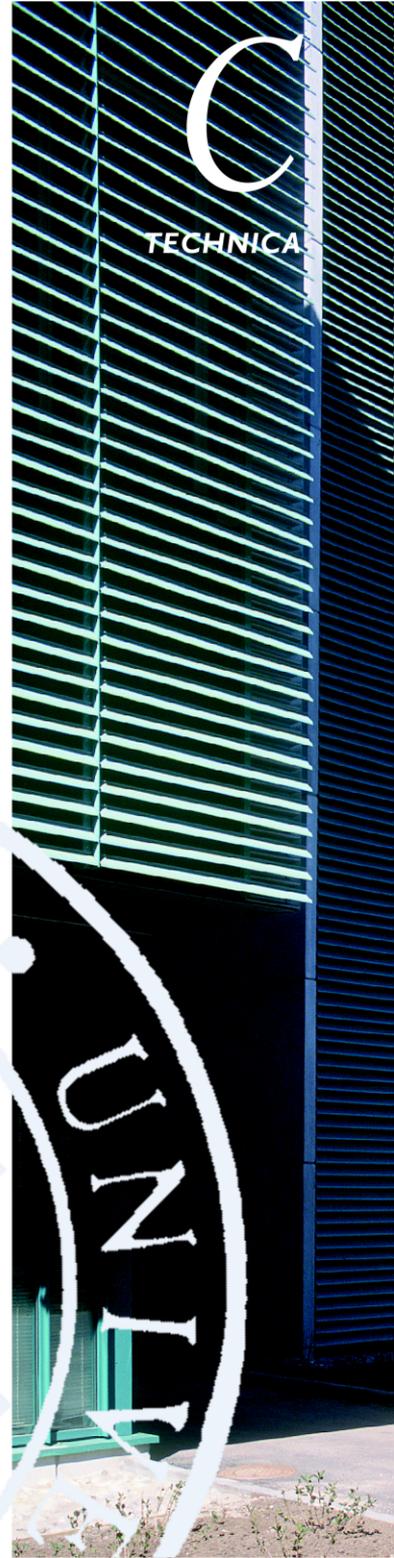

*ACTA*

*UNIVERSITATIS OULUENSIS*

C
TECHNICA

*Jonas Oppenlaender*

# CROWDSOURCING CREATIVE WORK

*JONAS OPPENLAENDER*

# CROWDSOURCING CREATIVE WORK

Supervised by
Associate Professor Simo Hosio

Reviewed by
Professor Florian Alt
Assistant Professor Ujwal Gadiraju

Opponent
Associate Professor Juho Kim

Cover Design
Raimo Ahonen

**Oppenlaender, Jonas, Crowdsourcing creative work.**
University of Oulu Graduate School; University of Oulu, Faculty of Information Technology
and Electrical Engineering
*Acta Univ. Oul. C 794, 2021*
University of Oulu, P.O. Box 8000, FI-90014 University of Oulu, Finland

### *Abstract*

Creative work is launched on paid crowdsourcing platforms, yet we lack an in-depth understanding of how the two key stakeholders of crowdsourcing platforms (crowd workers and requesters) perceive and experience creative work.

Creativity is a human characteristic that is difficult to automate by machines, and supplying requesters with crowdsourced human insights and complex creative work is, therefore, a timely topic for research. According to value-sensitive design, the integration of human insight into complex socio-technical systems will need to consider the perspectives of the two key stakeholders.

This article-based doctoral thesis explores the stakeholder perspectives and experiences of crowdsourced creative work on two of the leading crowdsourcing platforms.

The thesis has two parts. In the first part, we explore creative work from the perspective of the crowd worker. In the second part, we explore and study the requester's perspective in different contexts and several case studies.

The research is exploratory and we contribute empirical insights using survey-based and artefact-based approaches common in the field of Human-Computer Interaction (HCI). In the former approach, we explore the key issues that may limit creative work on paid crowdsourcing platforms. In the latter approach, we create computational artefacts to elicit authentic experiences from both crowd workers and requesters of crowdsourced creative work.

The thesis contributes a classification of crowd workers into five archetypal profiles, based on the crowd workers' demographics, disposition, and preferences for creative work. We propose a three-part classification of creative work on crowdsourcing platforms: creative tasks, creativity tests, and creativity judgements (also referred to as creative feedback). The thesis further investigates the emerging research topic of how requesters can be supported in interpreting and evaluating complex creative work.

Last, we discuss the design implications for research and practice and contribute a vision of creative work on future crowdsourcing platforms with the aim of empowering crowd workers and fostering an ecosystem around tailored platforms for creative microwork.

*Keywords:* creative work, creativity, creativity support tools, crowdsourcing

## Tiivistelmä

Luovassa työssä käytetään maksullisia joukkoistusalustoja, mutta meiltä puuttuu kuitenkin vielä syvällinen käsitys siitä, miten kaksi avainasemassa olevaa joukkoistusalustojen sidosryhmää (joukkotyöntekijät ja toimeksiantajat) ymmärtävät ja kokevat luovan työn. Luovuus on ihmisen ominaisuus, jota on vaikea automatisoida, ja joukkoistettujen inhimillisten näkemysten ja kompleksisen luovan työn välittäminen toimeksiantajille ovat siitä syystä ajankohtainen tutkimuskohde. Arvosensitiivisen suunnittelun mukaan inhimillisen ymmärryksen integroinnissa kompleksisiin sosioteknisiin järjestelmiin on otettava huomioon kahden avainasemassa olevan sidosryhmän näkökulmat.

Tässä artikkeliväitöskirjassa tutkitaan sidosryhmien näkökulmia ja kokemuksia joukkoistetusta luovasta työstä kahdella joukkoistusalustalla. Väitöskirja koostuu kahdesta osasta. Ensimmäisessä osassa tarkastellaan luovaa työtä joukkotyöntekijän näkökulmasta. Toisessa osassa tarkastellaan toimeksiantajan näkökulmaa useissa tapaustutkimuksissa. Tällä tutkimuksella halutaan syventää empiiristä ymmärrystä hyödyntämällä kyselytutkimuksiin perustuvia lähestymistapoja ja ihmisen ja tietokoneen välisessä vuorovaikutuksessa (Human-Computer Interaction, HCI) yleisiä lähestymistapoja. Ensimmäisessä lähestymistavassa tarkastellaan keskeisiä seikkoja, jotka voivat rajoittaa maksullisilla joukkoistusalustoilla tehtävää luovaa työtä. Jälkimmäisessä lähestymistavassa luodaan laskennallisia artefakteja, joilla halutaan tuoda esiin joukkotyöntekijöiden ja joukkoistetun luovan työn toimeksiantajien aitoja kokemuksia. Väitöskirjassa joukkotyöntekijät luokitellaan viiteen arkkityyppiprofiiliin, jotka perustuvat joukkotyöntekijöiden demografisiin tietoihin, ajattelumalleihin ja luovaa työtä koskeviin mieltymyksiin.

Väitöskirjassa ehdotetaan kolmiosaista luokittelua joukkoistusalustoilla tehtävälle luovalle työlle: luovuutta edellyttävät tehtävät, luovuustestit ja luovuuden arvioinnit (joita kutsutaan myös luovuuspalautteeksi). Lisäksi väitöskirjassa tutkitaan uutta tutkimusaihetta eli sitä, miten toimeksiantajia voidaan tukea monimutkaisen luovan työn tulkinnassa ja arvioinnissa. Lopuksi tutkimuksessa tarkastellaan mallin vaikutusta tutkimukseen ja käytäntöön, ja siinä esitetään tulevaisuuden joukkoistusalustoilla tehtävästä luovasta työstä visio, jonka päämääränä on parantaa joukkotyöntekijöiden valmiuksia ja tukea luovan mikrotyön räätälöityjen ympäristöjen ympärille rakentuvaa ekosysteemiä.

*Asiasanat:* joukkoistaminen, luova työ, luovuus, luovuutta tukevat työkalut

# Acknowledgements

The research presented in this thesis was conducted at the Center for Ubiquitous Computing at the University of Oulu, Finland.

I would like to thank Prof. Timo Ojala for generously funding my doctoral position at the Center for Ubiquitous Computing, and for giving me the exceptional opportunity to pursue my research interests independent of any research projects.

I would like to thank my supervisor and kettlebell supplier, Associate Prof. Simo Hosio, for his many deliveries of pink kettlebells, and Denzil Ferreira (2nd supervisor).

I thank the opponent, Associate Professor Juho Kim from the Korea Advanced Institute of Science and Technology (KAIST), and the two external reviewers, Prof. Florian Alt from Bundeswehr University Munich and Assistant Prof. Ujwal Gadiraju from Delft University of Technology. I would further like to thank the members of my Follow-up Group: Adjunct Prof. and Research Director Susanna Pirttikangas and Associate Prof. Georgi V. Georgiev.

I thank the co-authors who contributed to my first-authored papers during my PhD: Prof. Panos Ipeirotis, Associate Prof. Andrés Lucero, Associate Prof. Thanassis Tiropanis, Associate Prof. Halil Erhan, Dr. Jorge Goncalves, Dr. Aku Visuri, Dr. Naghmi Shireen, Dr. Abderrahmane Khiat, Dr. Maja Vuković, Kristy Milland, Elina Kuosmanen, Kennedy Opoku Asare, Maximilian Mackeprang, and Jesse Josua Benjamin.

The doctoral position was overall supported by the 6G Finnish Flagship Programme (Academy of Finland #318927) and was later continued as part of the University of Oulu's GenZ strategic profiling project (Academy of Finland #318930).

I thank the Nokia Foundation, the Finnish Foundation for Technology, the Jenny and Antti Wihuri Foundation, the Tauno Tönning Foundation, and the Riitta and Jorma J. Takasen Foundation for generously supporting me with their prestigious scholarships and grants. I further thank ACM SIGCHI for awarding me the SIGCHI Student Travel Grant to attend the Creativity & Cognition conference in 2019. Last, I would like to thank the UbiComp/ISWC 2018 Scholarship Committee for awarding me the UbiComp/ISWC student travel grant in 2018, and the University of Oulu Graduate School (UniOGS) for supporting my travels and studies with multiple grants.

August 11, 2021                                    Jonas Oppenlaender

# List of abbreviations

AAAI      Association for the Advancement of Artificial Intelligence

ACM      Association for Computing Machinery

AI      Artificial Intelligence

APA      American Psychological Association

AUT      Alternative Uses test

cf.      compare

CHI      CHI Conference on Human Factors in Computing Systems

e.g.      exempli gratia

et al.      et alia

ex-ante      before the event

HCC      Human-Centred Computing

HCI      Human-Computer Interaction

HIT      Human Intelligence Task

inter alia      among other things

IEEE      Institute of Electrical and Electronics Engineers

i.e.      id est

ML      Machine Learning

MTurk      Amazon Mechanical Turk

post-hoc      after this

UI      User Interface

# List of original publications

In the summary, references are cited by their Roman numerals:

I     Oppenlaender, J., Milland, K., Visuri, A., Ipeirotis, P., & Hosio, S. (2020) Creativity on Paid Crowdsourced Platforms. *Proceedings of the 2020 CHI Conference on Human Factors in Computing Systems (CHI '20)*, ACM: New York, NY, USA. 1–14. doi: 10.1145/3313831.3376677

II    Oppenlaender, J. & Hosio, S. (2019) Design Recommendations for Augmenting Creative Tasks with Computational Priming. *Proceedings of the 18th International Conference on Mobile and Ubiquitous Multimedia (MUM '19)*, ACM: New York, NY, USA. 35:1–35:13. doi: 10.1145/3365610.3365621

III   Oppenlaender, J., Kuosmanen, E., Lucero, A., & Hosio, S. (2021) Hardhats and Bungaloos: Comparing Crowdsourced Design Feedback with Peer Design Feedback in the Classroom. *Proceedings of the 2021 CHI Conference on Human Factors in Computing Systems (CHI '21)*, ACM: New York, NY, USA. 1–14. doi: 10.1145/3411764.3445380

IV   Oppenlaender, J. & Hosio, S. (2019) Towards Eliciting Feedback for Artworks on Public Displays. *Proceedings of the 2019 ACM Conference on Creativity & Cognition (C&C '19)*, ACM, New York, NY, USA. 562–569. doi: 10.1145/3325480.3326583

V    Oppenlaender, J., Kuosmanen, E., Goncalves, J., & Hosio, S. (2019) Search Support for Exploratory Writing. Lamas, D., Loizides, F., Nacke, L., Petrie, H., Winckler, M., & Zaphiris, P. (Eds.) *Human-Computer Interaction – INTERACT 2019 (LNCS 11748)*, Springer: Cham, Switzerland. 314–336. doi: 10.1007/978-3-030-29387-1_18

VI   Oppenlaender, J., Tiropanis, T., & Hosio, S. (2020) CrowdUI: Supporting Web Design with the Crowd. *Proceedings of the ACM on Human-Computer Interaction (PACMHCI '20)*. Vol. 4, No. EICS, ACM: New York, NY, USA. 76:1–76:28. doi: 10.1145/3394978

**Article I:** The thesis author designed the survey study, gathered and prepared the data, analysed the data, and led the writing.

**Article II:** The thesis author created and deployed a web-based survey instrument, designed the study, conducted the interviews, analysed the data, and led the writing.

**Article III:** The thesis author designed the survey study, analysed the data, and led the writing. The quantitative analysis was supported by the second author.

**Article IV:** The thesis author designed the study, conducted the interviews, analysed the data, and led the writing.

**Article V:** The thesis author designed, implemented, and deployed the software, designed and conducted the studies, collected and analysed the data, and led the writing.

**Article VI:** The thesis author developed the situated feedback system, designed the study, conducted the interview with artists supported by the co-author, conducted interviews with feedback providers, analysed the data, and led the writing.

# Contents

# 1    Introduction

## 1.1    Motivation

Creativity is one human characteristic that is difficult to automate by machines, "a bastion of human dignity in an age where machines [...] seem to be taking over routine skilled activities and everyday thinking" (Cropley, 2011). While low-level creative work, such as writing natural-sounding articles, can be completed by machine learning models (Brown et al., 2020), the innate human ability for generating ideas and finding solutions to "wicked" problems (Rittel & Webber, 1973) will play an increasingly important role in the future. Creativity is needed in various domain-specific and domain-general contexts, including, but not limited to, graphic design, user interface design, writing, art, and ideation (i.e., the generation of ideas). Creativity contributes to economic growth, as well as social, scientific, and technological innovation, with a potentially massive impact on global competitiveness (Florida, 2012; Mitchell, Inouye, & Blumenthal, 2003).

Crowdsourcing has emerged as a potential source for eliciting original thought and creative work. In practice, crowdsourcing is routinely used in behavioural and creativity-directed scientific experiments, as well as for eliciting subjective, creative thoughts and feedback online. Some of the key issues of crowdsourcing creative work on online crowdsourcing platforms are predicated on the crowd workers' willingness and motivation to complete creative work. The crowd workers' willingness to contribute to creative tasks is often taken for granted by the requester of creative work. Workers may, however, not be in the "mood" for creative work (Qiu, Gadiraju, & Bozzon, 2020). The crowd worker may rush to "satisfice" (Simon, 1956) and – motivated by a desire to maximise the income – try to complete creative tasks as quickly as possible. The crowd worker may, therefore, not produce results of high quality and validity (Hamby & Taylor, 2016). Monetary pay may not be the optimal way to motivate creative work (Frey & Jegen, 2001; Law et al., 2016; Mao et al., 2013; Rogstadius et al., 2011; Shaw, Horton, & Chen, 2011). Creative tasks are idiosyncratic and subjective, and it is, therefore, difficult to assess the quality of creative work. Out of fear of rejection (Gadiraju & Demartini, 2019), some crowd workers may shy away from creative work.

The perspective of the crowd worker has drawn little attention in prior literature (Martin, Hanrahan, O'Neill, & Gupta, 2014; Silberman, Irani, & Ross, 2010), and the subjective work experience of crowd workers is only recently receiving more interest in the scientific literature (e.g., Deng, Joshi, and Galliers (2016); Y. Wang, Papangelis, Lykourentzou, et al. (2020)). Following the approach of value-sensitive design (Friedman, 1996; Friedman & Hendry, 2019), the integration of human intelligence and insight into computation processes will need to consider the perspectives of the key stakeholders on crowdsourcing platforms. It is imperative to develop an understanding of how crowd workers, as one of the key stakeholders of crowdsourcing platforms, perceive and experience creative work. Similarly, it is important to understand how crowdsourced creative work is perceived and experienced by the requester of creative work. Such insights may inform the design of online decision and support tools that rely on *crowd-powered creativity* (Oppenlaender, Mackeprang, et al., 2019; Oppenlaender, Shireen, et al., 2019).

## 1.2    Overview of the thesis

The thesis is structured in two parts that correspond to two key stakeholders perusing online crowdsourcing platforms: Crowd workers and requesters (depicted in Figure 1).

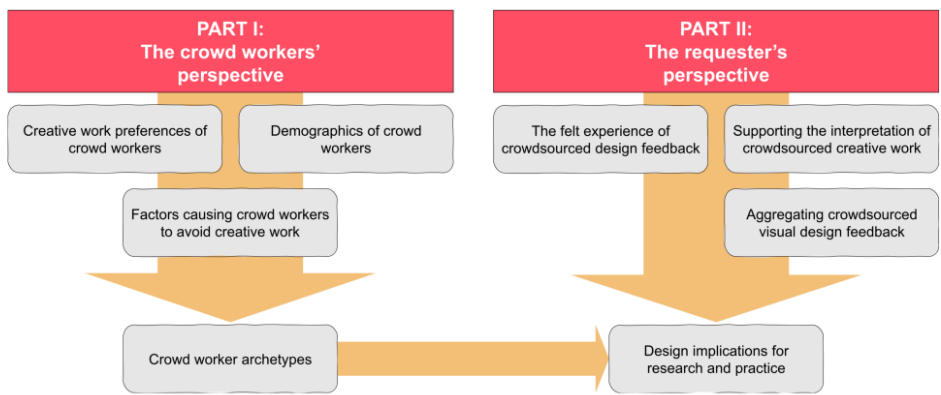

**Fig. 1. Overview of two perspectives covered by the thesis and summary of its contributions.**

The thesis includes six original articles which are referred to in the text by their Roman numerals (**I–VI**). Five articles have been published in relevant peer-reviewed, international conferences in the field of HCI (articles I–III, V, and VI). **Article IV** is a poster with preliminary results, presented at a relevant peer-reviewed, international conference.

**Article I** and **Article II** provide an investigation into the perspective of the crowd worker and computational priming, respectively. The logic progression of the other articles is as follows. **Article III** was designed for the specific purpose of studying the requester's felt experience of creative feedback. In **Article IV**, we created a situated feedback system with the intent of studying different mechanisms of collecting crowdsourced feedback. The investigation of both the requesters' and crowd workers' viewpoints revealed a gap between the needs of the requester and the crowd workers. This gap motivated the subsequent two articles. **Article V** studied how the requester can be supported in interpreting crowdsourced creative work supported by a faceted filtering interface. **Article VI** studied how the requester can be supported in making sense of complex creative work elicited from the crowd.

### 1.2.1    *Research objectives and questions*

The broad research aim of this thesis is to **explore stakeholder perspectives and experiences of creative work on crowdsourcing platforms**. We pursue this aim through two objectives:

1. gaining an understanding of the key issues that may limit creative work on general-purpose paid crowdsourcing platforms from the perspective of the crowd worker, and
2. creating computational artefacts to elicit authentic experiences from both crowd workers and requesters of crowdsourced creative work.

The research aim and objectives are supported by three research questions.

**RQ1.  How do crowd workers perceive creative work on the crowdsourcing platform?**

The first part of the thesis investigates creative work from the subjective viewpoint of the crowd worker, a viewpoint that has received little attention in prior literature (Martin et al., 2014; Silberman, Irani, & Ross, 2010). We study this research question on two complementary and commonly used crowdsourcing platforms and derive insights into the work preferences of different archetypes of crowd workers.

From the perspective of the requester of creative work, as the second key group of stakeholders on crowdsourcing platforms, we ask:

**RQ2. How do requesters perceive and experience creative work?**

Requesters face a number of challenges when evaluating and interpreting the crowd-sourced creative work. For instance, crowdsourcing platforms offer the potential to elicit vast amounts of diverse and potentially contradicting creative work and feedback (Yen, Kim, & Bailey, 2020). Requesters of creative work must identify the best solutions from among the mass of crowdsourced creative work and requesters may become overwhelmed with this task (Yen et al., 2020). In the last research question, we ask:

**RQ3. How can we support requesters in evaluating crowdsourced creative work?**

We investigate RQ2 and RQ3 in four different contexts (articles III–VI) in which we provide the requester with specific examples of complex creative work (articles V and VI) and creative feedback (articles III and IV) from crowd workers.

### 1.2.2 Article contributions

The author developed three web-based systems (articles IV–VI) and one web-based interface (Article II) to elicit authentic experiences from both crowd workers and requesters of crowdsourced creative work. Table 1 provides a summary of the four technical contributions. Articles I and III were survey studies and are not listed in Table 1.

**Table 1. Technical contributions of the thesis.**

| Article | Acronym | Description |
| --- | --- | --- |
| II | – | Web-based survey instrument to subject participants to computational priming |
| IV | SIMPLEX | Situated crowd feedback system for eliciting summative feedback on public displays |
| V | GAS | Adaptation of the web-based AnswerBot decision support system by Hosio et al. (2016) to support exploratory writing |
| VI | CrowdUI | Web-based system to support web design by eliciting visual design feedback from the crowd |

A summary of how each article contributes to the three research questions is provided in Table 2. The individual research contributions of each original article in this thesis are the following:

**Table 2. Contribution of the articles to the research questions. The main contributing articles for each research question are highlighted in bold.**

| Research questions | Articles |
|---|---|
| RQ1: How do crowd workers perceive creative work on the crowdsourcing platform? | **I**, II |
| RQ2: How do requesters perceive and experience creative work? | **III**, IV, V, VI |
| RQ3: How can we support requesters in evaluating crowdsourced creative work? | II, IV, **V**, **VI** |

**Article I** explores the crowd workers' perspectives on creative work on two general-purpose paid crowdsourcing platforms. The article contributes five archetypal profiles of crowd workers (Professional, Casual, Pragmatic, Novelty Seeker, and Self-developer), based on different perceptions and attitudes towards creative work. The article further identifies differences between the crowd workers of the two crowdsourcing platforms in both preferences and prior exposure to creative work, and identifies factors that may cause resentment against creative work in crowd workers. The article further contributes recommendations for requesters of creative work on general-purpose paid crowdsourcing platforms to avoid common pitfalls.

**Article II** evaluates computational priming as a task design method to explore whether roles can be used to make crowd workers more creative. The article further investigates three strategies of assigning roles to workers to overcome an impasse in the flow of ideas. The article finds that roles are not a silver bullet for making people more creative, but roles may be useful when a person reaches an impasse in the flow of ideas. Of three given strategies, self-selecting a single role was found to be the best strategy to overcome an impasse in the flow of ideas. The article contributes recommendations for the design of creative tasks with computational priming.

**Article III** explores how crowdsourced design feedback is experienced and perceived by students in the context of an HCI design course, and how crowdsourced design feedback compares to peer design feedback. The results of a mixed-method data analysis contribute an in-depth investigation of subjective factors (e.g., perceived usefulness and perceived fairness) of crowdsourced design feedback, as well as qualitative insights into how the students, as the feedback receivers, value crowdsourced design feedback.

**Article IV** provides a small-scale case study on crowdsourcing summative situated feedback on public displays. The author designed and evaluated a situated feedback system called *SIMPLEX*, and the article explored the potential of this system for supporting creative individuals with crowdsourced design feedback for digital artefacts. The interviews with the study participants focused on the evaluation of eight different types of design feedback, and the article contributes a discussion of potential problems when evaluating large-scale crowdsourced feedback. The study brought into view numerous challenges that requesters face when interpreting crowdsourced creative work. This inspired the subsequent line of inquiry in articles V and VI.

**Article V** presents a case study on supporting requesters of creative work in the exploration and selection of crowdsourced ideas. The system adapted for the purpose of this study (referred to as *GAS*) successfully demonstrated to support the exploration of crowdsourced viewpoints and ideas that best fulfil multiple criteria simultaneously. The article further contributes a writer-oriented understanding of the distinct search and exploration strategies and composing processes that people adopt when exploring information for a writing piece.

**Article VI** presents a case study of crowdsourcing visual design feedback for web design. The article further investigates an aggregation method to support the requester with evaluating the crowdsourced visual design feedback. The main contribution of the article is *CrowdUI*, a web-based system to elicit visual design suggestions from website users. The article established the feasibility of participatory visual design feedback and confirmed that visual feedback aggregation enables the requester of crowdsourced visual design feedback to draw actionable conclusions.

### 1.2.3    Structure of the thesis

The remainder of the thesis is organised as follows. Chapter 2 provides an overview on creativity and crowdsourcing and defines our understanding of 'creative work' on crowdsourcing platforms to clearly delineate the scope of the thesis. This is followed by an overview of the six articles included in this thesis and a summary of the research methods (Chapter 3). In Chapter 4, we explore the perspective of the crowd worker (RQ1). Chapter 5 explores the requester's perspective (RQ2 and RQ3). Chapter 6 revisits the research questions, discusses the findings and its limitations, and provides implications for future research and practice. Chapter 7 concludes the thesis.

# 2 Background and related research

This chapter introduces fundamental concepts from the related scientific literature that are used throughout the thesis. The second aim of this chapter is to clearly delineate the scope of the thesis in regard to crowdsourcing, creativity, and creative work.

## 2.1 Crowdsourcing

Crowdsourcing is the practice of outsourcing tasks to a crowd of people via an open call for contributions (Estellés-Arolas & de Guevara, 2012; Hosseini, Shahri, Phalp, Taylor, & Ali, 2015; Howe, 2006). The four main pillars of crowdsourcing are the task, the crowd worker (i.e., the person completing the crowdsourced task), the requester of the crowdsourced task, and the online platform on which the task is published (cf. Ali, Hosseini, Phalp, and Taylor (2014)). Regarding the motivation of workers to complete tasks, crowdsourcing can come in a variety of forms, e.g., compensation-based, contest-based, and volunteer or community-based crowdsourcing.

A major dichotomy used to define crowdsourcing is the distinction between microtask and macrotask crowdsourcing. In microtask crowdsourcing, complex tasks are decomposed into smaller units of work. On microtask crowdsourcing platforms, *requesters* launch short tasks (microtasks) for an anonymous crowd of *workers* to complete in exchange for a small monetary reward. Tasks on this type of crowdsourcing platform "are typically small, independent, homogeneous, have minor incentives, and do not require longer engagement from workers" (Roy, Lykourentzou, Thirumuruganathan, Amer-Yahia, & Das, 2013). Amazon Mechanical Turk[1] (MTurk) is, perhaps, the most popular platform for paid microtask crowdsourcing. On macrotask crowdsourcing platforms, on the other hand, tasks are more complex. A common assumption is that macrotasks take more time to complete, and likely require expert skills and knowledge to complete (Lykourentzou, Khan, Papangelis, & Markopoulos, 2019). The bulk of tasks on commercial crowdsourcing platforms are microtasks (Lykourentzou et al., 2019).

Another type of crowdsourcing platforms are online human subject pools, such as Prolific[2]. Prolific aims to provide an alternative to traditional means of recruiting participants, such as panels and human subject pools at universities. The work launched on Prolific is different from the microtasks that form the bulk of work on MTurk. Tasks on Prolific predominantly consist of online surveys and the time to complete a task is longer than a few seconds. As such, Prolific is a suitable complement to analyse creativity on online crowdsourcing platforms.

---

[1]https://www.mturk.com

[2]https://www.prolific.co

Crowdsourcing is a powerful method for tapping into the collective insights of a crowd of people with a diverse background and knowledge (Surowiecki, 2005). Crowdsourcing has become part of the toolkit of many researchers as a means for recruiting study participants and conducting online studies and experiments (Behrend, Sharek, Meade, & Wiebe, 2011; J. Chandler & Shapiro, 2016; Horton, Rand, & Zeckhauser, 2011; Kittur, Chi, & Suh, 2008; Paolacci, Chandler, & Ipeirotis, 2010; Stewart, Chandler, & Paolacci, 2017), and the interest in using crowdsourcing platforms for academic studies and experiments has been growing (Anderson et al., 2019). Researchers have developed a good understanding of the demographics of crowd workers on crowdsourcing platforms (e.g., Buhrmester, Kwang, and Gosling (2011); Difallah, Filatova, and Ipeirotis (2018); Hara et al. (2018); Ipeirotis (2010a, 2010b); Martin et al. (2014); Paolacci et al. (2010); Ross, Irani, Silberman, Zaldivar, and Tomlinson (2010); Stewart et al. (2015)), and a large body of work explored, for instance,

– how workflows can be devised and optimised (e.g., Dow, Kulkarni, Klemmer, and Hartmann (2012); Kucherbaev, Daniel, Tranquillini, and Marchese (2016)),
– how complex work can be orchestrated (e.g., Retelny et al. (2014); Salehi and Bernstein (2018); Vaish et al. (2017); Valentine et al. (2017)),
– how work results can be improved by framing tasks with different contexts (e.g., August and Reinecke (2019); Lewis, Dontcheva, and Gerber (2011); Morris, Dontcheva, and Gerber (2012); Teevan and Yu (2017)),
– how the cost and completion times of crowdsourced tasks can be modelled and minimised (e.g., Difallah, Catasta, Demartini, and Cudré-Mauroux (2014); Faridani, Hartmann, and Ipeirotis (2011); Gao and Parameswaran (2014); Karger, Oh, and Shah (2014); Minder, Seuken, Bernstein, and Zollinger (2012); Singer and Mittal (2011, 2013)), and
– how the quality of crowdsourced data can be improved with various quality control measures (see Daniel, Kucherbaev, Cappiello, Benatallah, and Allahbakhsh (2018) for an overview).

### 2.1.1    *Prior work comparing crowdsourcing platforms*

An important consideration when devising a crowdsourcing campaign is the selection of the right crowdsourcing platform, and whether data should be collected on a crowdsourcing platform in the first place. A number of researchers have compared different participant pools and crowdsourcing platforms.

Paolacci et al. (2010) and Kees, Berry, Burton, and Sheehan (2017) compared MTurk with a traditional subject pool, such as students at a university. The latter is a popular means of convenience sampling for experiments and studies in HCI. Paolacci et al. (2010) investigated the trade-offs of different methods of recruiting participants and concluded that participants recruited on MTurk are a viable alternative compared to participants from traditional student subject pools and online discussion boards. Kees et al. (2017) found that data collected on MTurk was comparable to data collected with panels and student subject pools. MTurk workers performed particularly well in common methods of testing attention and concentration, such as instructional manipulation checks (Oppenheimer, Meyvis, & Davidenko, 2009). Bentley, O'Neill, Quehl, and Lottridge (2020) compared ten different means of recruiting participants common in HCI (e.g., survey panels, MTurk, and employees) and found that MTurk was superior in completeness, thoughtfulness, and grammatical correctness. The authors concluded by recommending MTurk for the collection of open-ended data.

Horton et al. (2011) found the behaviour of crowd workers on MTurk to be consistent with the behaviour observed in traditional laboratory experiments. Komarov, Reinecke, and Gajos (2013) also found no significant differences between online experiments on MTurk and laboratory experiments. Gadiraju et al. (2017) suggested using crowds from MTurk as a complement to laboratory experiments. Mason and Suri (2012) recommended MTurk as a useful tool for conducting behavioural research online. This indicates that crowdsourcing platforms may be a conducive setting for conducting user studies, such as the unsupervised evaluation of user interfaces.

Other researchers, however, were more cautious in recommending MTurk for user studies and for collecting data. Kittur et al. (2008), Ahler, Roush, and Sood (2019), and Gadiraju, Kawase, Dietze, and Demartini (2015) found that a large percentage of data collected on MTurk may be potentially untrustworthy. Kittur et al. (2008), for instance, found that crowd workers from MTurk may provide uninformative, semantically empty, and non-constructive responses in subjective microtasks. To combat the "widespread

'gaming' of the system," Kittur et al. recommended adding verifiable questions to subjective tasks.

Peer, Brandimarte, Samat, and Acquisti (2017) compared a traditional participant pool, Prolific, MTurk, and another crowdsourcing platform. The authors found demographic differences between the crowdsourcing platforms. For example, participants on Prolific spent the least amount of weekly hours working on their crowdsourcing platform (and consequently earned less money). Prolific participants demonstrated the highest degree of naïveté among the participants from the three crowdsourcing platforms. Prolific participants also demonstrated "a lower propensity [...] to engage in dishonest behaviors, as compared to MTurk" (Peer et al., 2017).

Labour market demographics, of course, change with time (Silberman, 2015). But another key to understanding the vast differences in the studies reviewed in this section lies in getting a better understanding of the qualitative work experience of crowd workers.

### 2.1.2    Prior work on the perspective of the crowd worker

Much of the research on crowdsourcing is conducted with primarily the requester's benefits in mind (Martin et al., 2014) and with a focus on incremental improvements in work and data quality (Vakharia & Lease, 2015). The requester-centred focus of academia largely fails to consider the subjective perspective and work reality of the crowd workers as one of the main stakeholder groups on online crowdsourcing platforms. It is, therefore, not unfounded to ask what crowd workers think about non-routine cognitive work, such as creative work. However, qualitative studies focusing entirely on the perspective of the crowd worker are rare in the literature.

First, one line of inquiry to better understand the perspective of the crowd workers is content analysis. Crowd workers can be considered as networked individuals who emphatically share their feelings and concerns in communities online (Gray, Suri, Ali, & Kulkarni, 2016; Yin, Gray, Suri, & Vaughan, 2016). Crowd workers, therefore, leave traces on online message boards, such as Turker Nation[3]. The content on these online fora can be linguistically analysed. For instance, Nouri, Wachsmuth, and Engels (2020) mined different online forums for problems reported by crowd workers. Inter alia, the authors reported on problems in the evaluation of tasks (e.g., unfair rejections) related to the prevalent power dynamics between the requesters and crowd workers on paid crowdsourcing platforms. Martin et al. (2014) analysed the content of Turker Nation. Among other findings, the authors reported on the crowd workers' view of MTurk and the relationship between crowd workers and requesters.

---

[3]https://www.reddit.com/r/TurkerNation/

Second, the subjective experience of crowd workers can be elicited in online surveys. Kittur et al. (2013), for instance, collected the perspective of a sample of crowd workers on MTurk for their article on the future of crowd work. The "bill of rights" (Irani & Silberman, 2013) and the work by Whiting et al. (2017) on assembling a constitution for crowdsourcing marketplaces are further examples of worker-centred investigations of crowdsourcing. Silberman, Irani, and Ross studied the problems of crowd workers on MTurk from a worker's point of view (Silberman, Irani, & Ross, 2010; Silberman, Ross, Irani, & Tomlinson, 2010). More recently, Y. Wang, Papangelis, Lykourentzou, et al. (2020) launched a survey study on a Chinese crowdsourcing platform to analyse the subjective work experiences of crowd workers. The authors' results are in line with prior findings "that the work experiences of crowdworkers [...] were affected by stressful deadlines and complex requirements in tasks" and "dehumanizing crowdwork conditions" (Y. Wang, Papangelis, Lykourentzou, et al., 2020). Y. Wang, Papangelis, Saker, et al. (2020) also investigated the experiences of crowd workers in China. The authors found significant differences in the work context, motivation, engagement, and preferences between individual crowd workers and crowd workers in so-called crowd farms.

Last, the voice of the crowd worker can be collected via tool-supported means. Turkopticon by Irani and Silberman (2013) is an example of a worker-centred collective system that allows crowd workers to rate and review HITs and requesters. The tool collects the subjective work experiences of crowd workers. Reviews are, however, often shaped by the work reality on the microtask platform (e.g., abusive requesters, unfairly rejected work, or underpayment), and rarely mention creative work.

## 2.2 Creativity

In this section, we first provide a brief background on creativity to frame our subsequent definition of creative work and our investigation of creative work on crowdsourcing platforms. The main aim of this section is to demonstrate the diversity of different ways of defining and measuring 'creativity' and to provide context for the remainder of the thesis.

'Creativity' is a multi-faceted concept that is difficult to define, as evident by the extensive literature (e.g., Boden (1996); Kaufman and Sternberg (2010); Runco (2014); Runco and Pritzker (2011); Sawyer (2012); Sternberg (2005); Weisberg (2006)). In the field of Human-Computer Interaction, the diversity of literature and different definitions of creativity have led to a certain "fuzziness" with which the term 'creativity' is used in research (Frich, MacDonald Vermeulen, Remy, Biskjaer, & Dalsgaard, 2019; Frich, Mose Biskjaer, & Dalsgaard, 2018). In the remainder of this section, we review a number of classifications and dichotomies from the scientific literature to frame the concept of 'creativity.'

By and large, creativity can be viewed from a *domain-specific* and a *domain-general* perspective. While, as the name suggests, domain-specific creativity is specific to a given domain (e.g., music and architecture), domain-general creativity is a transferable skill that can be applied in a broad number of different contexts. Creativity can be studied through a historical lens ("*H-creativity*") or on a personal, psychological level ("*P-creativity*") (Boden, 1991, 1996). The four C model of creativity by Kaufman and Beghetto (2009) directly relates to this dichotomy. The four C model comprises four different levels of creativity: *Big-C*, *Pro-C*, *mini-c*, and *little-c*. The latter two can be summarised as *small-c* creativity. *Big-C* creativity is also referred to as *eminent* creativity (Richards, 1990). Big-C creativity makes a long-lasting contribution to society. For instance, the inventions of the polymath Leonardo da Vinci and the compositions of Wolfgang Amadeus Mozart fall under the umbrella of Big-C creativity. Big-C creativity is typically studied through a historical lens by analysing biographies of genius-level inventors and eminent figures in history (Simonton, 2013). *Small-c* creativity, on the other hand, refers to the creativity of ordinary persons who create something on a personally meaningful level (Richards, 1990). For instance, this could include a person discovering a novel method of learning (*mini-c*) or the creation of a painting that is valuable for a limited group of people, such as family members (*little-c*) (Kaufman & Beghetto, 2009).

Another four-component lens on creativity is the "four P" model of creativity by Rhodes (1961): *person*, *process*, *product*, and *"press." 'Person'* relates to characteristics and behaviour inherent to a person, including the person's intellect, traits, habits, and personality. Historically, this is where the research community's attention rested in the early days of research on creativity since creativity was understood, at the time, as a concept that is correlated with intellect (Guilford, 1950, 1956, 1959). *'Process'* refers to the processes that make a person creative, such as perception and thinking. *'Press'* refers to the external environment that may affect a person's creativity. *'Product'* refers to the embodiment of an idea in a tangible form. A product "presents a record of a [human]'s thinking at some point in time" (Rhodes, 1961), and creativity, under this view, can therefore be studied by analysing the creative products of a person.

Creativity is regarded as a social construct in the literature (Amabile, 1983a; Amabile, Goldfarb, & Brackfleld, 1990), and the contemporary research community commonly refers to the traditional paradigm of the lone inventor as a "myth" (Lemley, 2012). Studies from psychology show that groups of people with diverse backgrounds provide high-quality ideas and can outperform skilled experts (Hong & Page, 2004). Investigations into collaborative creativity dominate the field of creativity research in HCI and form the "second wave" of research on creativity (Frich et al., 2019, 2018).

Many authors regard creativity as a process with a number of distinct phases. A classic example of a personal model of creativity is Wallas' model of the creative process with four phases: preparation, incubation (maturing), illumination (creation of insights and inspiration), and verification (execution) (Wallas, 1926). In an applied context in design and engineering, process models typically go through subsequent divergent and convergent phases to open and narrow down the solution space, respectively. The view of creativity as a process is particularly relevant in the domain of design where processes typically also iteratively go through divergent and convergent phases – see, for example, the double diamond process model proposed by the British Design Council Design Council (2007) and the five-phase Design Thinking process taught by the Hasso-Plattner Institute of Design at Stanford (Plattner, Meinel, & Leifer, 2011; Simon, 1996).

Zeng, Proctor, and Salvendy (2011) abstracted the different creative processes found in the creativity-directed literature in a general model of the creative process that consists of *ideation*, *analysis*, *implementation*, and *evaluation*. In the field of HCI, Frich et al. (2019) synthesised a similar six-part process of creativity from the literature on creativity support tools. In the process by Frich et al., *pre-ideation* refers to any activities related to planning and management of the creative activity. *Idea generation or ideation* is the divergent activity of generating ideas. *Evaluation or critique* refers to the convergent activity of selecting ideas from a pool of generated ideas. *Implementation* relates to implementing the idea, validating, and verifying the implemented idea. The process steps are typically *iterated* in practice. *Meta or project management* refers to activities necessary to plan and launch creative work.

The difficulties to define the elusive concept of creativity also transcend into research on measuring creativity, as outlined in the following section.

## *Measuring creativity*

Historically, the research community's interest in measuring creativity was kickstarted by Guilford's presidential address to the American Psychological Association (APA) in 1950 (Guilford, 1950). This started the first wave of creativity research, which was strongly shaped by the prior efforts of measuring individual human intelligence (Frich et al., 2018; Runco, 2014). Later efforts in the 1980s (the second wave) shifted to the study of collaboration and the production of collaborative products as a measure of creativity (Frich et al., 2018; Sawyer & DeZutter, 2009).

When it comes to measuring the creativity of artefacts and ideas, the criteria used to measure creativity in practice are diverse. Some of the most common ways of measuring the creativity of an artefact or an idea include, but are not limited to, the

– **rarity of the idea**, measured, for instance, with the constructs ***originality*** (e.g., Dontcheva, Gerber, and Lewis (2011); Kerne et al. (2007); Kerne and Smith (2004); Lewis et al. (2011); Rietzschel, Nijstad, and Stroebe (2010); Runco and Jaeger (2012); Villalba Garcia (2008); Yu and Nickerson (2011, 2013)), ***novelty*** (e.g., Abdullah, Czerwinski, Mark, and Johns (2016); Andolina et al. (2017); Chan et al. (2017); Dennis, Minas, and Bhagwatwar (2013); Piffer (2012); Plucker, Beghetto, and Dow (2004); Poetz and Schreier (2012); Shireen et al. (2011); Siangliulue, Chan, Gajos, and Dow (2015); Teevan and Yu (2017); Yu, Kittur, and Kraut (2014, 2016); Zeng et al. (2011)), ***uniqueness*** (e.g., Ashktorab et al. (2020); Sethapakdi and McCann (2019)), ***unusualness*** (e.g., Tsai (2009)), ***non-obviousness*** (e.g., Tsai (2016)), or ***similarity*** (e.g., Shi et al. (2017)),

– **"fit"** with a certain audience, measured for instance with the constructs ***appropriateness*** (e.g., Abdullah et al. (2016); Kozbelt and Durmysheva (2007); Piffer (2012); Zeng et al. (2011)), ***usefulness*** (e.g., Buisine, Guegan, Barré, Segonds, and Aoussat (2016); Plucker et al. (2004); Tsai (2009); Yu et al. (2014)), ***effectiveness*** (e.g., Runco and Jaeger (2012)), ***"impact"*** (e.g., Piffer (2012)), ***practicality*** (e.g., Abbas, Khan, Tetteroo, and Markopoulos (2018); Kerne et al. (2007); Kerne and Smith (2004); Teevan and Yu (2017); Yu et al. (2014, 2016); Yu and Nickerson (2011, 2013)), ***feasibility*** (e.g., Lewis et al. (2011); Poetz and Schreier (2012); Rietzschel et al. (2010)), ***relevance*** (e.g., Dennis et al. (2013)), ***value*** and ***benefit*** (e.g., Andolina et al.

(2017); Poetz and Schreier (2012); Siangliulue, Chan, et al. (2015)), or *quality* (e.g., Kozbelt and Durmysheva (2007); Shireen et al. (2011)),

– **fluency** (i.e., the number of ideas generated) (e.g., Buisine et al. (2016); Chan et al. (2017); Kerne et al. (2007); Kerne and Smith (2004); Lewis et al. (2011); Villalba Garcia (2008)), or *quantity* (e.g., Andolina, Klouche, Cabral, Ruotsalo, and Jacucci (2015); Dennis et al. (2013); Shi et al. (2017); Shireen et al. (2011); Siangliulue, Chan, et al. (2015)),

– **flexibility** (i.e., the number of different categories of ideas) (e.g., Kerne et al. (2007); Kerne and Smith (2004); Lewis et al. (2011); Villalba Garcia (2008)), sometimes also expressed as *breadth* (e.g., Mackeprang, Khiat, and Müller-Birn (2018)), *diversity* (e.g., Chan et al. (2017)), or *variety* (e.g., Shireen et al. (2011)), or

– **elaboration** (i.e., the length of an idea) (e.g., Lewis et al. (2011); Villalba Garcia (2008)).

Other authors simply measure the *"creativeness"* (e.g., Ashktorab et al. (2020); Siangliulue, Arnold, Gajos, and Dow (2015); Tsai (2009)) or *"creativity"* (e.g., Buisine et al. (2016); Kozbelt and Durmysheva (2007)) of artefacts or ideas. Less common ways of measuring creativity include, for instance, *likeability* (Kozbelt & Durmysheva, 2007) and *funniness* (Ashktorab et al., 2020). Combinations of criteria can also be found in the literature, such as, for instance, *"appropriate novelty"* (Abdullah et al., 2016) and *"creative usefulness"* (Mann & Cadman, 2014).

The above list of creativity metrics is based on the author's review and perusal of the literature over three years of doctoral studies. While the list is by no means a representative and complete survey of the literature, it highlights the "criterion problem" endemic to much of the literature (Amabile, 1983b). The literature's use of measurement instruments in the field of HCI is as fragmented as the literature on creativity itself (c.f. Frich et al. (2018)). In practice in the field of HCI, approaches to measuring creativity are chosen based on their fit with the study design and their power to contribute to a sensible explanation of empirical phenomena.

## 2.3    Creative work

From the perspective of the crowd worker on general-purpose paid crowdsourcing platforms, we divide the relevant literature into three categories (see Figure 2). The categories are the three primary ways in which crowd workers encounter *creative work* on crowdsourcing platforms.

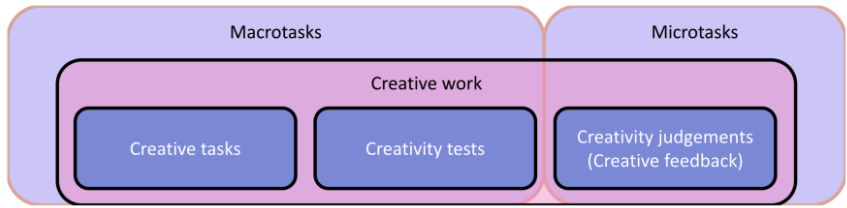

**Fig. 2. Creative work on general-purpose paid crowdsourcing platforms.**

In a **creative task**, the crowd worker is asked to contribute ideas, creative artefacts, or provide solutions to a given problem, following the paradigm of Open Innovation (Baldwin & von Hippel, 2011). Creative tasks can manifest in a variety of different ways, such as, for instance, *ideation*, that is, the generation of new ideas (e.g., Siangliulue, Arnold, et al. (2015); Siangliulue, Chan, et al. (2015)), *drawing* and *sketching* (e.g., Cai, Jongejan, and Holbrook (2019); Doan, Ramakrishnan, and Halevy (2011); Koblin (2009)), *creative writing* (e.g., Bernstein et al. (2010); C.-Y. Huang, Huang, and Huang (2020); Kim, Sterman, Cohen, and Bernstein (2017); Kittur, Smus, Khamkar, and Kraut (2011); Salehi, Teevan, Iqbal, and Kamar (2017)), and *problem-solving* (e.g., Cranshaw and Kittur (2011); Lakhani, Fayard, Levina, and Pokrywa (2012); Lakhani and Lonstein (2011)).

**Creativity tests** are standardised means to directly measure an individual's creativity. The context of this type of creative work is, typically, an online study that aims to measure creativity as the dependent variable in the experiment design. Crowd workers are asked to complete a standard test of divergent thinking (such as the Alternate Uses test (Guilford, Christensen, Merrifield, & Wilson, 1978) or the Torrance Tests of Creative Thinking (Torrance, 1966)), convergent thinking (such as the Remote Associates test by Mednick (1962)), or puzzles and practical insight problems (such as the candle problem by Duncker (1945), the eight-coin problem (Öllinger, Jones, Faber, & Knoblich, 2013), or the two strings problem by Maier (1931)).

**Creativity judgements** or **creative feedback** are a type of creative work in which crowd workers are asked to assess the creativity of creative artefacts (Amabile, 1983b). While this type of creative work does not require the crowd worker to generate artefacts or ideas, it does involve assessing the creativity of artefacts, ideas, or other people. Judgements are often used to assess work outcomes by other crowd workers (e.g., by calculating a rating score and verifying inter-rater agreement). For example, creativity-directed studies often ask workers to assess the creativity of ideas (e.g., Lykourentzou, Ahmed, Papastathis, Sadien, and Papangelis (2018)). Another example of this type of creative work is formative and summative design feedback. Summative design feedback is geared towards evaluating the final output of creative work, while formative design feedback is intended to provide pointers for improving a work-in-progress (Sadler, 1989).

The similarity between the two former types of creative work is rooted in the crowd workers' involvement in the creation of the artefact or idea. In creative tasks and creativity tests, the crowd worker generatively applies creativity and leaves "a unique thumbprint on the output" (Oppenheim, 2005). The following two examples demonstrate that from the perspective of the crowd worker, creative tasks and creativity tests can be similar in their felt experience:

> Example 1a: *"Come up with birthday messages for Mary, a firefighter"* (Siangliulue, Arnold, et al., 2015)
> Example 1b: *"Think of unique and unusual uses for a paperclip"* (Lewis et al., 2011)
>
> Example 2a: In the Sheep Market project by Koblin (2009), crowd workers from Amazon Mechanical Turk were asked to "draw a sheep facing left".
> Example 2b: In the "Invented Alien Creature" test (Ward, 1994), participants are asked to imagine and draw a life-form from another planet.

Example 1a is an instance of a creative task, while Example 1b is an instance of a standardised creativity test (the Alternate Uses Test by Guilford et al. (1978)). Both examples 1a and 1b ask the worker to generate short written texts until the worker produces a sufficient amount of examples or runs out of ideas. Common measures of creativity, such as originality, feasibility, elaboration, flexibility, and fluency (Guilford, 1967) can be applied to determine the quality of the work results. In Example 2, workers express their creativity in sketches. While the results of Example 2a were used to assemble a crowdsourced artwork, Example 2b can be considered as a test of creativity (Kozbelt & Durmysheva, 2007). Again, both examples are similar in the skills required to complete the task and how the task is experienced by the person completing the task.

In practice, creative work is launched on general-purpose paid crowdsourcing platforms for instance in creativity-directed studies, but also in two types of crowdsourcing systems described in the following section.

## 2.4    Crowdsourcing creative work

Crowds are heterogeneous and offer a diverse set of skills (Paolacci & Chandler, 2014; Surowiecki, 2005). Organisations such as Innocentive, Quirky, and OpenIDEO successfully apply the concept of Open Innovation (Baldwin & von Hippel, 2011) by crowdsourcing diverse ideas and leveraging the different contexts and backgrounds of crowd workers (Dennis & Williams, 2003). Crowdsourcing is especially effective in situations that require human cognition for decision-making, such as complex creative work (Kittur et al., 2011). In the remainder of this section, we elaborate on creativity support tools and crowd feedback systems as two ways how creative work is crowdsourced in the practice of HCI research.

### *Creativity support tools*

The goal of creativity support tools (Shneiderman, 2002, 2007, 2009; Shneiderman et al., 2006) is to make "more people more creative more often" (Shneiderman, 2002, 2007). Supporting creativity is considered a grand challenge of HCI (Shneiderman, 2009; Stephanidis et al., 2019), and research on supporting creativity builds on a long line of work on studying human creativity (Guilford, 1950) and augmenting human intellect (Engelbart, 1962). The NSF Workshop on Creativity Support Tools (Shneiderman et al., 2006) can be considered as a pivotal event for research into creativity support tools. Interest in creativity support has since been growing in the field of Human-Computer Interaction, and Frich et al. identified 143 creativity support tools in their review of HCI literature (Frich et al., 2019).

Humans excel in creative work, such as recombination, analogical transfer, and divergent thinking. It, therefore, makes sense to involve humans in creativity support tools. Shneiderman (2002) proposed eight activities that could benefit from being supported by creativity support tools: searching, visualising, consulting peers, thinking, exploring, composing, reviewing, and disseminating. Crowd-powered systems (Bernstein, 2012) and crowd-powered creativity support tools (Oppenlaender, 2020; Oppenlaender, Mackeprang, et al., 2019; Oppenlaender, Shireen, et al., 2019) have the potential to support these eight activities by drawing on a diverse crowd (Surowiecki, 2005). Such systems also have the potential to facilitate access to creative work and could create new opportunities for creative work in regions of the world that previously did not have access to the international labour market (Paritosh, Ipeirotis, Cooper, & Suri, 2011; Shneiderman, 2002).

### Crowd feedback systems

Crowd feedback systems (Luther et al., 2015; Xu, Huang, & Bailey, 2014) are a specific type of crowd-powered systems to asynchronously collect feedback, decision support, and critique from a large number of people online (Kou & Gray, 2017). The potential of collecting feedback from the crowd has been explored and investigated in studies about feedback and a number of crowd feedback systems. It is impossible to account for all literature here, but we briefly review some instances of crowd feedback systems in the remainder of this section to exemplify this line of research.

*Voyant* by Xu et al. (2014) is a system that draws on a non-expert crowd to elicit feedback on graphic designs. The system focuses on capturing the first impressions and how well a graphic design meets its stated goals. *CrowdCrit* by Luther et al. (2015) provides designers with formative feedback on graphic designs also written by non-expert crowd workers. *Paragon* by Kang, Amoako, Sengupta, and Dow (2018) complements written feedback on graphic designs with visual design examples that helped the crowd to provide feedback that was more specific and actionable. *Decipher* by Yen et al. (2020) aims to support the requester with interpreting design feedback. The system aggregates and visualises the sentiment of a collection of feedback. The visualisation helped the users of the system to "feel less overwhelmed during feedback interpretation tasks" (Yen et al., 2020).

## 2.5    Scope of the thesis

Given the vast array of research on creativity, crowdsourcing, and creativity support in HCI, it is imperative to clearly define the scope of this thesis and clarify the author's use of terminology. This section explains the scope of the thesis in regard to creativity, crowdsourcing, and creative work. Following Borning and Muller (2012), this section is also provided as a form of self-disclosure to clarify the thesis author's assumptions and epistemological standpoint.

The scope of this thesis, in regard to creativity, is as follows (see summarised in Figure 3). First, as opposed to computational creativity which seeks to simulate or replicate human creativity in machines (Colton & Wiggins, 2012), the research in this thesis is rooted in human-centred computing (HCC) and places humans at the centre of research (Guzdial, 2013). As such, the thesis is scoped to human creativity. Second, creativity is viewed as a domain-general concept in this thesis, as opposed to creativity that requires domain-specific knowledge. Third, we argue that creativity at the highest, paradigm-shifting level (Big-C) is not likely to be found in data collected on microtask crowdsourcing platforms. While a case could be made that some crowd workers operate on the Pro-C level of creativity – for instance, workers who professionally pursue creative writing tasks – the majority of contributions on microtask crowdsourcing platforms can be expected to be from "non-experts" who are motivated by small-c creativity. Fourth, in the four P model by Rhodes (1961), we take a product-based view of creativity. Under this view, a creative activity has a tangible or intangible outcome that is of value to at least one person or a wider group of people, e.g. an idea, a sketch, or a feedback item. Fifth, our product-based focus on creative outcomes shapes our choice of measuring creativity. We primarily adopt one of the most widely used definitions of creativity in the literature (Runco & Jaeger, 2012). Following this definition, a product (or an idea) must be novel (original, unique, or unusual) and useful (effective, valuable, or appropriate) to be considered creative. Last, we choose to make individuals our unit of study, not groups. We root our choice to study individuals in communication theory (Shannon & Weaver, 1949). Information is always first processed by a single mind (the individual receiving the information) before it is made available to and shared in a group of people. As such, our choice to factor out collaborative creativity is a pragmatic first step towards future investigations into collaborative creativity support.

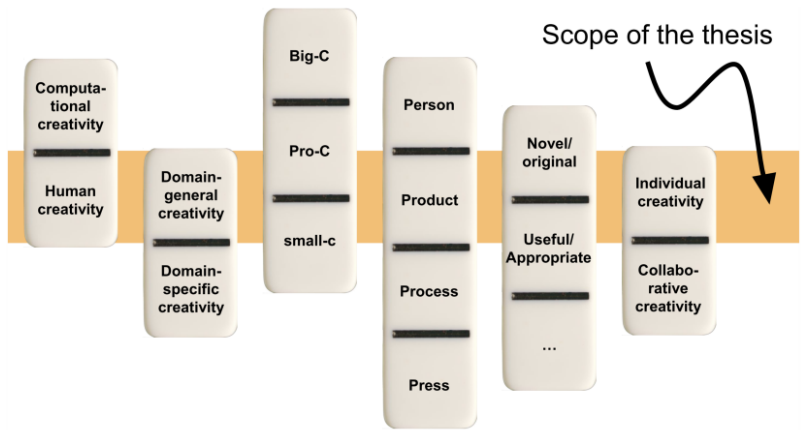

**Fig. 3. Graphical summary of the scope of the thesis in regard to creativity.**

### Scope in crowdsourcing

Creative work can be found on several types of crowdsourcing platforms online. First, there are crowdsourcing platforms created for the specific purpose of collecting ideas and innovative solutions, such as communities for online feedback exchange (Foong, Dow, Bailey, & Gerber, 2017) and websites for ideation (e.g., Innocentive, Quirky, OpenIDEO, Tricider, and Synthetron). Second, there are different variants of crowdsourcing for eliciting human insights, such as citizen science and design contests. Last, there are online marketplaces for expert services, such as digital work (e.g., Upwork, Freelancer, Topcoder) and design (e.g., 99designs, CAD Crowd, jovoto).

This thesis focuses on crowdsourcing platforms on which creative work is rare as of yet. We refer to these crowdsourcing platforms as *general-purpose paid crowdsourcing platforms* (henceforth just crowdsourcing platforms). Crowd workers on these crowdsourcing platforms are anonymous and typically complete "human intelligence tasks" (HITs). Since the tasks require human insight, they are typically domain-general and do not require expert or domain-specific knowledge. The tasks are typically small, repeatable, and offer only a small monetary reward per piece (microtasks). As such, this type of crowdsourcing work is akin to piecework in a factory (Alkhatib, Bernstein, & Levi, 2017). Creative work on these crowdsourcing platforms is launched, but underexplored. Studying creativity on these crowdsourcing platforms, therefore, presents an opportunity where this thesis can make a timely contribution.

The thesis primarily focuses on two of the leading general-purpose paid crowdsourcing platforms: Amazon Mechanical Turk[4] and Prolific[5]. Amazon Mechanical Turk was created for outsourcing small microtasks, while Prolific provides a means for recruiting study participants for empirical online surveys and experiments. The two platforms are different in work scope and, therefore, provide complementary insights.

---

[4] www.mturk.com

[5] www.prolific.co

Regarding the type of crowd, the thesis primarily focuses on paid crowdsourcing. However, the thesis also includes one study that was conducted with volunteers (community members). The choice of crowd was mandated by the study material and the purpose of the tool that was being evaluated. Since the study was conducted about a specific website and the tool's purpose was to improve the said website, we expected the participants to have some knowledge of the website and intrinsic motivation to complete the task. Also, this task was, unlike the other tasks, more complex and required longer times to complete (macrotask).

### Scope in creative work

In this thesis, we study creative tasks, creativity tests, and creative feedback as the three primary ways of how creative work is requested on general-purpose paid crowdsourcing platforms. We also use creative judgements in some of our own articles as a means to evaluate crowdsourced data. Table 3 summarises what type of creativity is studied (e.g., as the dependent variable in the study design) in the six different articles.

**Table 3. Types of creative work studied and applied in this thesis.**

| Type of creative work | Studied in articles |
| --- | --- |
| Creative tasks | I, V, VI |
| Creativity tests | I, II |
| Creativity judgements | III, IV, VI |

### The thesis in the context of the requester's creative process

The majority of creativity support tools focus on the early ideation phase of the creative process (Frich et al., 2019; K. Wang & Nickerson, 2017). The research in this thesis is primarily situated at the end of the requester's creative process. That is, we are interested in supplying creative work to the requester and supporting the requester in evaluating this creative work. The requester, in our case, can be thought of as the user of a crowd-powered creativity support tool or crowd feedback system who requests creative work from the crowd.

# 3    Research methods

The work in this thesis was conducted by using a mix of quantitative and qualitative research methods common in the field of HCI. This section provides a brief overview of the research methods used in the different articles.

## 3.1    Research approach

Supporting creativity with technology is inherently an exploratory and trans-disciplinary endeavour (Mitchell et al., 2003). Research in this space must draw on methods from different scientific disciplines, such as, but not limited to, Sociology, Psychology, and Computer Science. As an interdisciplinary research field, Human-Computer Interaction (HCI) touches on all these disciplines (Lazar, Feng, & Hochheiser, 2017) and acts as an umbrella for this thesis. The research in this thesis is further founded in human-centred computing (Guzdial, 2013) and value-sensitive design (Friedman, 1996; Friedman & Hendry, 2019), and "places humans [...] at the center of the research" (Guzdial, 2013). The research is exploratory and investigates the perspective of the two key stakeholder groups on crowdsourcing platforms.

Overall, the research follows a pluralistic approach. We make both empirical contributions and artefact contributions. Our exploration draws on mixed methods (Tashakkori & Creswell, 2007) and seeks to explore creative work on crowdsourcing platforms in multiple different contexts and with the following research methods.

## 3.2    Experimental methods

### *Laboratory experiments*

Articles IV and V employed controlled experiments in a laboratory setting. Laboratory experiments are one way of obtaining behavioural empirical data on the usability of software (Wolf, Carroll, Landauer, John, & Whiteside, 1989). As such, we used laboratory experiments to evaluate a software artefact and to study participants using the software's user interface. In the case of Article IV, the controlled environment consisted of a workplace in an office with a desk, chair, and a laptop computer which allowed the participant to use and evaluate a web-based user interface under the observation of a researcher. In Article V, the controlled environment was provided in the form of an installation with a public display and a tablet device in a secluded space on the university campus. Participants were invited to use the installation and were observed by a researcher. Both laboratory studies were designed as within-subject experiments in which participants used different user interfaces.

### Online experiments

In articles II and VI, participants completed tasks in their own natural environment without supervision by a researcher. For Article II, a custom web-based survey instrument was developed and implemented to subject crowd workers from an online crowdsourcing platform to different conditions in an online within-subject experiment. Article VI used a remote study setup in which participants were asked to use and evaluate a web-based software artefact on their own desktop computers. Participants completed a pre and post task questionnaire integrated into the software artefact. A number of interaction metrics were logged in this study, such as the number of times a participant interacted with an element, and the time spent using the software artefact.

### Extension of prior studies

Articles II and III shared ground with prior studies and employed study setups similar to prior literature. Replication studies hold value in the field HCI (Hornbæk, Sander, Bargas-Avila, & Grue Simonsen, 2014), and the replication setup allowed us to validate and expand upon prior studies. Article II investigated the prior finding that computational priming may positively affect the outcome of creative work by crowd workers (Lewis et al., 2011; Morris et al., 2012; Teevan & Yu, 2017). Article III extended prior studies by Dow, Gerber, and Wong (2013), Hui, Glenn, Jue, Gerber, and Dow (2015), and Wauck et al. (2017), particularly concerning the requester's perception and felt experience of crowdsourced feedback.

## 3.3    Data collection

The two key methods for eliciting information from study participants in this thesis were online surveys and interviews. Throughout this thesis, quotes from participants and crowd workers are highlighted in italics.

*Online survey studies*

All articles (I – VI) elicited information from participants in online questionnaires. In the case of articles I and III, online survey studies were the one and only means of eliciting data from participants. In all other articles, the online questionnaires were supplemented with data collected via qualitative methods, such as interviews. The online questionnaires included both open-ended items and quantitative items (typically on a 7-point anchored Likert scale).

*Interviews*

In articles II, IV, and V, interviews with participants were conducted as a follow-up and complement to the structured collection of data in online questionnaires. Article II used interviews to gain insights into the participants' experiences during the study and on creativity in general. The interviews, in this case, were conducted as dyadic interviews which allowed us to profit from the social dynamics and interaction between the two participants (Morgan, Ataie, Carder, & Hoffman, 2013). In Article IV, we used interviews to explore the first impressions, the benefit and usefulness of our system, as well as the needs of one potential user group. Information from another group of potential users was collected in unstructured interviews following a laboratory study. In Article V, semi-structured one-on-one interviews were conducted immediately after the participant filled the questionnaire. The interviews focused on observations during the laboratory study and the participants subjective experience of using the user interface.

## 3.4    Data analysis

The analysis of the data was conducted with a mix of quantitative and qualitative methods. When quantitative analysis was needed, we used quantitative methods typical in the field of HCI. Adequate statistical tests were used, such as parametric tests (e.g., t-test and CHI-square) and non-parametric tests (e.g., Wilcoxon rank sum). The qualitative analysis was methodologically rooted in grounded theory (Glaser & Strauss, 1967). Thematic analysis was used for developing insights from the qualitative data (Braun & Clarke, 2006). Article III applied rigorous content analysis.

### 3.5    Research ethics

All related ethical procedures were followed as required by the thesis author's university. The author and his co-authors were engaged to follow good scientific practice in accordance with the Declaration of Helsinki, the national ethical guidelines of Finland, and the European Charter for Researchers. The research was committed to the overriding obligations of research: openness, quality and accountability (Norwegian National Committee for Research Ethics in Science and Technology, 2016). Code, data, and other study-related material was made available as auxiliary material of the articles or in online code repositories.

The research engaged people online or offline as study participants. All related ethical procedures for human-subject experiments were followed as required by the University of Oulu. In each study, participants were asked to provide consent prior to participating in the study, and participants consented to their data being used (in anonymised form) for the purpose of the academic studies. Participants were also recruited on crowdsourcing platforms (Prolific and Amazon Mechanical Turk). The participating crowd workers were compensated according to the principles of fair crowd work (Silberman et al., 2018).

# 4  The worker's perspective on creative work

In studies in HCI and related fields, crowd workers are routinely asked to contribute to ideation by completing creative work in crowdsourcing campaigns. However, the willingness of the crowd on paid crowdsourcing platforms to participate in creative work is rarely examined. In this first part of the thesis, we explore and investigate creative work on crowdsourcing platforms from the viewpoint of the crowd worker.

## 4.1  Creative work preferences of crowd workers

The demographics of crowd workers may provide an insight into the work reality on crowdsourcing platforms. Based on a sample of 215 crowd workers in our survey study in Article I, we found differences in demographics between the crowd workers from Amazon Mechanical Turk and Prolific. Crowd workers on MTurk were more likely to work professionally and worked longer hours on the crowdsourcing platform, compared to crowd workers on Prolific. "Professional crowdworkers" (Gleibs, 2017) or "super turkers" (Bohannon, 2011; Toxtli, Richmond-Fuller, & Savage, 2020) have been shown to complete the bulk of work on their crowdsourcing platform. This type of worker is likely to be found in samples on MTurk, as "80% of the tasks are carried out by the 20% most active" crowd workers (Martin et al., 2014). Our survey study in Article I confirmed this impression. Workers on MTurk were also more likely to have a long work history on MTurk, compared to crowd workers on Prolific. Crowd workers on Prolific were a more casual workforce, earned less of their monthly income on the crowdsourcing platform, and spent fewer weekly work hours on their crowdsourcing platform, compared to crowd workers on MTurk (see Figure 4).

Through the lens of the crowd workers, creative work was found to be rare on both crowdsourcing platforms. This, and the difference in demographics of the two crowdsourcing platforms (as described above), confirmed our belief that the two crowdsourcing platforms form a fruitful and complementary ground for studying creativity in a context where creativity is not the norm.

Crowd workers on Prolific were more enthusiastic about and open towards creative work compared to crowd workers on MTurk (see Figure 5). Crowd workers on Prolific were more engaged than the MTurk workers and thought the tasks were entertaining, inspiring, and thought-provoking to complete. To the crowd workers on Prolific, tasks

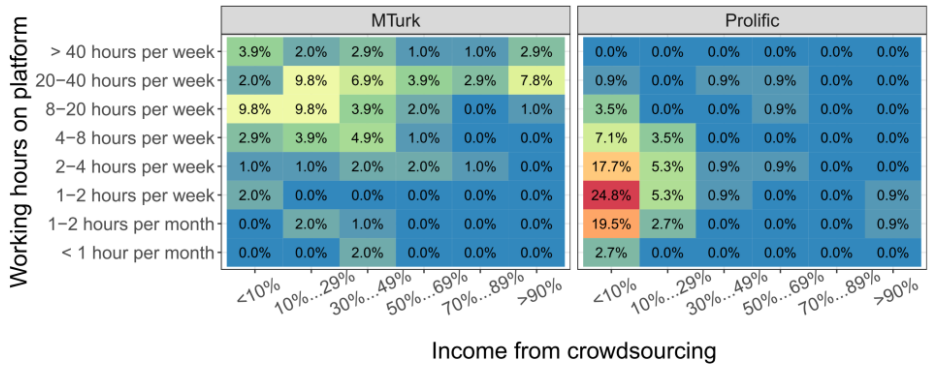

Income from crowdsourcing

**Fig. 4. Crowd worker responses to two multiple-choice questions: working hours on the respective crowdsourcing platform versus percentage of the worker's income from crowdsourcing (rounded to one decimal; reprinted with permission from Article I © ACM, 2020).**

that involve using one's imagination, such as creative writing, were seen as a *"fruitful," "stimulating,"* and *"fun"* way to spend one's free time and express oneself. Crowd workers on Prolific reported not having seen as much creative work as crowd workers on MTurk, and crowd workers on Prolific wanted to see more creative work on their platform.

Overall, crowd workers on Prolific preferred creative work because it allows a degree of freedom of expression in the work itself. However, crowd workers may also have reservations against creative work. In the remainder of this section, we identify factors that may cause crowd workers to avoid creative work.

## 4.2    Factors causing crowd workers to avoid creative work

*Monetary reward for creative work*

Crowd workers are primarily extrinsically motivated, especially on MTurk (Rogstadius et al., 2011). MTurk workers are under pressure to complete tasks quickly and strive to maximise their income (Gadiraju, Kawase, & Dietze, 2014). Since creative work is subjective, it may be associated with more uncertainty in rewards in the mind of the crowd workers.

Based on the findings in Article I, a clear concern for the crowd workers was underpayment for creative work. Creative work may take more time to complete than regular microtask work, and this may make the work seem to be *"too long and*

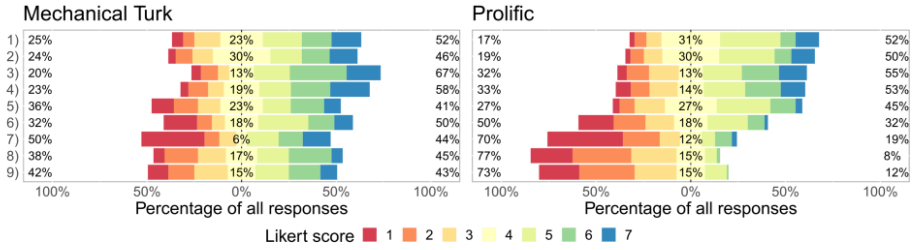

1) How many creative tasks should there be on <platform>? (1: Much less, 7: Much more)

2) How many creativity tests should there be on <platform>? (1: Much less, 7: Much more)

3) Overall, the creative tasks on <platform> you participated in were... (1: Not at all interesting, 7: Extremely interesting)

4) Overall, the creativity tests on <platform> you participated in were... (1: Not at all interesting, 7: Extremely interesting)

5) What type of tasks do you generally prefer on <platform>? (1: Simple and easy tasks, 7: More complex tasks that make me think)

6) Looking at your past participation in creativity studies on <platform>, how much do you think you learned about yourself? (1: Nothing at all, 7: Extremely much)

7) "I first learned about creativity tests on <platform>." (1: Strongly Disagree, 7: Strongly Agree)

8) How often have you seen creative tasks being offered on <platform>? (1: Not at all often, 7: Extremely often)

9) How often have you seen creativity tests being offered on <platform>? (1: Not at all often, 7: Extremely often)

**Fig. 5. Responses from crowd workers on various perceptions of creative work. The percentages on the left, middle and right indicate disagreement (1–3), neutrality (4) and agreement (5–7), respectively. For example, 77% of the Prolific workers indicated that they did not see creative tasks often on their crowdsourcing platform, while 8% encountered them more frequently (reprinted with permission from Article I © ACM, 2020).**

*complicated"* in the eyes of the crowd workers. Professional workers, in particular, may avoid creative work to maximise their income. Crowd workers expressed their dissatisfaction with the monetary rewards for creative tasks, for instance in comments such as

> *"creative tasks do not pay enough and exclude too many people on too many tasks. There are better platforms for these sorts of things."*

Other crowd workers mentioned that creative work is *"just not worth the time or effort for the money they offer"* and that – even though the crowd workers would like to see more creative work on the crowdsourcing platform – *"the money comes from the boring hits."*

### Unfair rejections of subjective creative work

Creative tasks are open-ended and subjective. While some crowd workers with a more casual approach to work welcomed these properties, professional crowd workers seek to avoid uncertainty in rewards. Unfair rejections of subjective work are a factor that may cause a crowd worker to shy away from creative work. One MTurk worker, for instance, reported that

> *"creative work is fun, but I only do creative or subjective work for requesters who are known to be fair requesters – I have suffered unfair rejections just because a requester didn't like my answers."*

Another MTurk worker urged requesters to *"not reject work for subjective assignments unless it is clear and unequivocal that the worker was scamming the requester."*

### Prior exposure to creative work

"Over-surveying" is a concern on crowdsourcing platforms (Behrend et al., 2011). Foreknowledge of and exposure to standardised tests, such as creativity tests, could negatively affect the robustness, reliability, effect size, and validity of creativity studies (J. Chandler, Mueller, & Paolacci, 2014; J. Chandler, Paolacci, Peer, Mueller, & Ratliff, 2015; Haigh, 2016; Hamby & Taylor, 2016; Robinson, Rosenzweig, Moss, & Litman, 2019). According to results obtained in Article II, the attitude of a crowd worker towards creative work may affect how the creative work is perceived. Crowd workers with a negative initial attitude towards creative work may think that the task is more difficult than crowd workers with a positive initial attitude.

Our work in Article I provides empirical evidence that crowd workers may not be naïve to commonly used creativity tests, as a subset of creative work on crowdsourcing platforms. A significant number of workers in our sample recalled having completed instances of common creativity tests. In a sample of 215 crowd workers, 167 workers (77.7%) were able to accurately describe an encounter with creative work. One-third of these crowd workers described an instance of a standardised creativity test. This is a remarkably high proportion of crowd workers, since workers were prompted to describe a single instance of creative work on the crowdsourcing platform, and the creativity test was first to come to mind. Most likely this is an underestimation of the true proportion, as crowd workers might have participated in creativity tests but did not recall an instance in our survey. Guilford's Alternative Uses Test (AUT), a test for divergent thinking, was particularly well known among the crowd workers ($N = 36$). The AUT is a test that, according to the results in Article II, may be perceived by crowd workers as a moderately difficult and moderately challenging task that requires a high level of concentration.

Over-surveying and the non-naïveté of crowd workers may contribute to workers resenting creative work. One crowd worker from MTurk, for example, urged requesters to *"Stop repeating the task! I have answered the 'come up with creative uses for a brick' question about fifty times. Pick something else!!!"* In the same vein, other participants noted that some tasks given to them have a *"classic textbook character,"* and remarked that *"People who do lots and lots of surveys and work on MTurk a lot will end up seeing these over and over, thus maybe invalidating any value the platform offers to requesters."*

### Feeling overwhelmed with creative work

Participants in Article VI reported feeling *"overwhelmed"* with the creative task. Beginner-level crowd workers, in particular, reported that they felt *"intimidated"* by the freedom afforded by the creative task. One participant, for instance, mentioned that *"with limited interface knowledge, I was a bit intimidated 'poking' around."*

This feeling was shared by the participants in Article IV. Creative work may have surprising properties that may confuse the crowd workers. When presented with more complex ways of collecting creative feedback, such as video and camera "selfies," participants mentioned that they simply did not know how to respond and what to do. This confusion, however, was also predicated by the surprising properties of the specific feedback collection mechanism which were given to the participant without prior explanations on how the technology was intended to be used. Three participants, for instance, used the video feedback to verbally address the feedback requester directly by explaining their thoughts.

### Concerns about the intellectual property of creative work

On a platform in which creative work is not the norm, creative work may be perceived as *"exploitative"* in that it may seem, in the eye of the crowd worker, more appropriate for websites and services that were created for the specific purpose of creative work. As one crowd worker from Prolific put it, *"creative tasks seem more appropriate for sites/jobs that pay their authors/artists/creative workers an appropriate wage [...] It feels exploitative to use a survey site under the 'disguise' of research."* Other crowd workers raised concerns about a lack of transparency on how their creative work is being used, such as a crowd worker from MTurk: *"I feel that my creativity is my own and if I wish to use I will. I don't want to give my ideas to a person or company I don't know."*

### Preference for working alone

Since creativity-directed research in HCI focuses predominantly on collaborative aspects (Frich et al., 2018), we investigated the crowd workers' preference for working alone. Article I presents empirical evidence that crowd workers have a strong preference for working individually. Almost 90% of the 215 crowd workers surveyed in Article I preferred to work alone. Only a minority (10.7%) of the crowd workers preferred working collaboratively. Crowd workers were concerned about potential issues when cooperating with other workers. Idle time and delays could negatively affect the crowd workers' productivity and efficiency. One crowd worker was concerned about a potential mismatch in skills between the collaborators which could further cause delays. Among the minority of crowd workers who preferred to work collaboratively, some informally mentioned that they simply enjoy the company of other people, or prefer collaborative tasks because this type of task is exciting.

## 4.3 Crowd worker archetypes

Based on the findings of the survey study in Article I, we contribute a classification of workers into five archetypal profiles based on the crowd workers' attitudes and perception of creative work (summarised in Figure 6).

*Professional workers* are extrinsically motivated by monetary rewards. The professional worker is likely to work long hours, and effectiveness and productivity, therefore, matters for this type of crowd worker. The professional worker may shy away from creative work for two reasons. First, collaboration with other crowd workers could involve technical delays and unforeseeable issues with other crowd workers which may impact the crowd worker's productivity and bottom line. Second, creative work is subjective, and requesters may unfairly reject this kind of work in the mind of the professional worker.

*Casual workers* are more open towards creative work and prefer tasks that allow the use of imagination. Rejections of work are less critical for this worker, as the worker does not rely on the income from the crowdsourcing platform as the primary form of income. This type of crowd worker is also more open to collaboration and creative experiments. This was also evident in the responses from the beginner-level crowd workers in Article VI who were more open to experimenting with new tools. Casual workers are also more likely to have never encountered creative work on the crowdsourcing platform.

**Fig. 6. Crowd worker archetypes based on their perception and attitude towards creative work on general-purpose paid crowdsourcing platforms.**

*Novelty seekers* enjoy creative work on the crowdsourcing platform. This type of crowd worker seeks out new experiences, and creative work offers an interesting and exciting alternative to *"boring"* and repetitive tasks. This type of crowd worker is also more open towards collaborative creative work. Games and collaborative experiments provide an enjoyable pass-time for this kind of crowd worker.

*Self-developers* see the tasks on the crowdsourcing platform as an opportunity to learn about themselves and their own personality. This worker enjoys creative work because it challenges the worker and may cause the worker to self-reflect and gain new knowledge and skills. Some crowd workers under this archetype reported that creative work may either develop their creative skills, or at least help discover one's own creative ability, as exemplified by the comment from a crowd worker on Prolific: *"I learned that I have more problem solving skills than I give myself credit for."* This type of crowd worker is curious about the outcome of a study and appreciates a debriefing after the creative work is completed.

*Pragmatic workers* hold the view that the crowdsourcing platform is a marketplace in which requesters launch work for crowd workers to complete, and someone will eventually accept any kind of work, including creative work. As such, the pragmatic worker is ambivalent about creative work. Creative work is just another type of task in the view of this type of worker. To make an informed decision of whether to accept the creative work or not, it is important for this type of crowd worker to be provided an accurate description of the creative work.

# 5 The requester's perspective on crowdsourced creative work

This part of the thesis includes four case studies, each conducted in a different context: **Article III** explores how crowdsourced creative feedback is experienced and perceived by undergraduate students (as the requesters) in a design-oriented HCI course. **Article IV** explores creative feedback in a small-scale case study on crowdsourcing situated summative feedback on a public display. **Article V** presents a case study on supporting requesters of creative work in the post-hoc exploration and selection of crowdsourced ideas for a writing piece. **Article VI** presents a case study of crowdsourcing and aggregating visual design feedback to support web design. The remainder of this chapter briefly summarises and synthesises the overarching results of the four case studies to answer research questions RQ2 and RQ3.

## 5.1 The felt experience of crowdsourced creative feedback

In the following, we investigate creative feedback, as a part of creative work relevant to the formative and summative stages of the requester's creative process. The findings in this section are based on Article III.

### 5.1.1 Crowdsourced creative feedback versus peer feedback

We found significant differences between peer feedback and creative feedback from MTurk. Students perceived creative feedback from their peers as being significantly more specific, actionable, explanatory, relevant, and of higher quality. Peer feedback was significantly longer, and students showed significantly fewer signs of satisficing compared to crowd workers on MTurk. Overall, the students were significantly more satisfied with the peer feedback than with crowdsourced feedback.

Students preferred the open-ended feedback over Likert-scale feedback, although students mentioned that they appreciated being able to see a direct numerical comparison between the two different feedback sources. The numerical summary of the Likert-scale feedback also allowed the students to identify weaknesses and strengths that were common in the feedback from both feedback sources. For instance, a student mentioned that *"since both sources are not correlated, it was easy to identify the main design failures in the app and prioritise and solve them."*

### 5.1.2    Perceived quality of creative feedback

Crowdsourced creative feedback was overall perceived as being less meaningful, less comprehensive, less constructive, less specific, and less actionable than peer feedback. Creative feedback from MTurk workers was perceived as more generic than peer feedback, to a point where many students deemed the crowdsourced creative feedback to be nonsensical due to its shortness and irrelevance to the task. Consequently, the students thought the crowdsourced creative feedback was less useful for the given task of designing a mobile application.

When it comes to the valence of the creative feedback (i.e., the affective tone or "harshness" of the feedback), some students appreciated the crowdsourced creative feedback for being more lenient. Overall, the crowdsourced creative feedback from MTurk was perceived as being less harsh and critical than peer feedback. This prompted some students to think that *"the class noticed the bad sides and the online workers noticed the good sides of the app."*

A difference in the effort put into writing the creative feedback was noticed by most of the students. The level of elaboration affected the students' perception of the usefulness of the feedback. Several students mentioned that the creative feedback from MTurk workers was thin in takeaways. Students acknowledged that their peers had put more thought and effort into writing the creative feedback compared to the crowd workers from MTurk. As commented by one student, the *"online workers just tried to speedrun the questions."* Crowdsourced creative feedback was thus perceived as being less useful in helping with the design task.

### 5.1.3    Perceived fairness of creative feedback

Even though the crowdsourced creative feedback was perceived as being less critical and harsh, it was nevertheless perceived as being less fair than peer feedback by about two thirds of the students. Students experienced the fairness of the creative feedback primarily along five dimensions, as depicted in Figure 7.

*Agreeableness* refers to the perception of how well the feedback aligns with the student's own point of view. Feedback that lacked criticism was often found to be more agreeable and, thus, was perceived as being fairer. However, if the creative feedback was well justified and sensible, students felt it was easier to reconcile the differences in opinion and still perceived the feedback as fair. A minority group of the students found the crowdsourced creative feedback more agreeable because it surfaced fewer problems in the design.

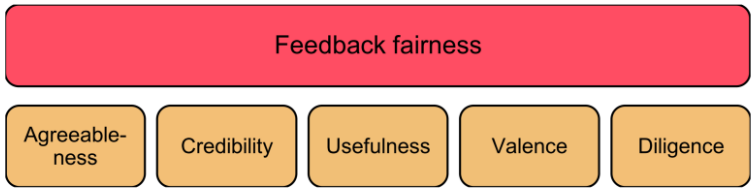

**Fig. 7. Students perceived fairness of creative feedback along five dimensions.**

The perceived *credibility* of the feedback provider strongly influenced the students' sense of feedback fairness. The fact that crowdsourced workers may provide an outsider's perspective inspired a minority of students to trust in the credibility of the crowd worker. The majority of the students, however, perceived the peer feedback as more realistic and better scoped to the course. The perceived credibility of the creative feedback from crowd workers was low in this group of students, as they realised that some of the creative feedback contained *"joke answers"* that were entered by crowd workers in order to satisfice the given task. Some students, however, admitted that their lack of faith in the feedback provider's credibility was partially motivated by simply not knowing who had given the feedback. Diversity of different viewpoints in a feedback item added to the feedback item's credibility and thus its perceived fairness.

The *usefulness* of the feedback was determined in part by its perceived helpfulness. Helpfulness, in the words of the students, constituted, for instance, suggestions, hints, concrete examples, and specific pointers on what to improve. *Useful* feedback, in general, was perceived as being fair. Perceived knowledge of the context (in this case, the HCI design course) also contributed positively to the students' sense of feedback fairness.

*Valence* refers to the affective tone and harshness of the feedback. Crowdsourced creative feedback was perceived as less critical, less harsh, and more positive in tone, compared to peer feedback. The more positive affective tone did, however, not necessarily contribute positively to the students' sense of fairness. The majority of students appreciated the more critical peer feedback for being fairer than crowdsourced creative feedback. The latter was often perceived to be overly positive, approving, and praiseful which contributed negatively to the sense of fairness in some students.

*Diligence* was perceived by the students along several dimensions. Elaborate replies were seen as being more diligent than short replies. Superficial feedback with a low specificity was perceived as less diligent. Whether the feedback was internally consistent and clear also contributed to the students' sense of diligence, and hence fairness.

### 5.1.4    Monetary valuation of the feedback

We identified two camps of students with different approaches to estimating the feedback's value in Article III. The "Hardhats" estimated the feedback's value in an engineer-like, strict, and grounded way. Given that members of this group thought that the feedback held little usefulness to them, the members of this group attributed no value to the crowdsourced feedback. The other group of students – which we call "Bungaloos" – had a more naïve and bona fide approach of estimating the feedback's value. Among the latter group, we found a mismatch between the perceived usefulness of the crowdsourced creative feedback and the students' monetary valuation of the feedback. Driven by conflated expectations and a mental image of the crowd worker as a professional usability expert, students in the Bungaloo group attributed value to crowdsourced creative feedback, even if the feedback was sometimes of little use to the students.

The bona fide trust in crowdsourced data was also noticeable in Article V. Study participants accepted information from a crowdsourced knowledge base as trustworthy because participants suspected that it consisted of a mix of non-specialist and specialist opinions. This mix was even considered to be more trustworthy than the results of a popular search engine. While participants could not articulate why they blindly trusted the crowdsourced knowledge base, one participant compared the information to Wikipedia, which – in the mind of this participant – contains trustworthy information. However, several participants did voice concerns about the provenance of the crowdsourced information which, due to the anonymous nature of the crowdsourcing model, could not be determined.

## 5.2    Supporting requesters with interpreting creative work

Creative feedback, and crowdsourced creative work in general, may provide requesters with valuable information. However, different types of crowdsourced creative work are not equally useful for the requester. Articles IV and VI, for instance, surfaced a potential problem for the requester of creative feedback: complex creative feedback cannot easily be interpreted at scale and does not aggregate well. While feedback providers preferred simple and quick-to-complete types of creative feedback, it was specifically the more complex creative feedback that holds rich information and insights for the requester.

In this section, we present three case studies in which the requester is supplied with complex creative work. We explore and investigate how the requester can be supported in interpreting this creative work (RQ3). Section 5.2.1 summarises the relevant findings of articles II and IV. Section 5.2.2 presents the relevant key findings of Article VI.

### 5.2.1    *Supporting the interpretation of crowdsourced creative work*

#### *Supporting requesters in the post-hoc exploration of crowdsourced data*

In Article IV, we designed and tested an intervention to support requesters in finding and exploring different ideas for writing an article. The intervention enables its users to explore the information of a crowdsourced knowledge base through a faceted filtering interface. The intervention provides decision support in form of short textual ideas that can be sorted according to multiple different criteria.

We conducted this research in the context of "exploratory writing" which we define as the written documentation of a complex online search, i.e., an online search that comprises at least two search criteria. A searcher may, for instance, search for *low-budget* movies that have *epic* special effects. The current search engines are ill-equipped for answering complex search questions because the search engines insufficiently enable the exploration of the "long tail" of search keywords (Bernstein, Teevan, Dumais, Liebling, & Horvitz, 2012). Instead, search engines optimise their search results by focusing on the most-searched single search keywords.

Our study compared the intervention to Google Search and found the intervention to be supportive in exploring an information space and selecting ideas. The study established the suitability of the intervention for supporting exploratory writing and information discovery. The users of the intervention reported being able to find an initial idea for the writing piece significantly more easily, compared to using both the tool and the search engine. the intervention facilitated the exploration of the information space, and significantly reduced the participant's mental effort. The intervention enabled its users to write articles that better matched the given criteria, and users reported that the articles were of higher quality overall.

### Narrowing the solution space with computational priming

Computational priming refers to the implicit manipulation of a task to induce behaviour changes (Lewis et al., 2011; Morris et al., 2012). Computational priming may be a potential way to *ex-ante* limit the solution space (before data is even collected) and, therefore, reduce the complexity of the requester's task of interpreting the crowdsourced creative work.

Based on our findings and experience in Article II, we conclude that roles are not a silver bullet for augmenting the creativity of crowd workers. Contrary to prior research, we could not find a significant difference in the "unusualness" of the ideas created under three different computational priming conditions. However, our findings highlight that adopting roles can be an effective tool for creative workers to overcome an impasse in the flow of ideas. Roles may further be useful as a means of narrowing down the solution space to support the requester in evaluating crowdsourced creative work. Responses from crowd workers on Prolific were more nuanced and scoped when primed with a role. Computational priming may be a suitable strategy to filter ideas and, therefore, support the requester in evaluating the crowdsourced ideas by preventing cognitive overload. Of the three given strategies, self-selecting a single role was found to be the best strategy to overcome an impasse in the flow of ideas.

### 5.2.2 Aggregating crowdsourced visual design feedback

In Article VI, we created an artefact to elicit visual design feedback from crowd workers. The artefact – called *CrowdUI* – is a tool for running remote experiments. Participants undergo multiple process steps in which they first enter demographic data about themselves in a tool-integrated questionnaire, then modify a given user interface, then evaluate one or multiple user interfaces created by other participants, and then fill out another concluding tool-integrated questionnaire. The tool was created with the intent of supporting the requester in the complex creative work of designing a website.

The tool was evaluated in two requester-centred studies. The intention of the first study was to explore the feasibility of crowdsourcing visual design feedback from non-experts (in this case, community members). The second study explored issues that surfaced in the first study when web developers attempted to evaluate the crowdsourced user interfaces.

In the first study, web developers were presented with a representative sample of crowdsourced user interfaces. In the eyes of the web developers, the tool provided a perspective on the needs of users and *"concrete ideas and opinions from users as actual design."* Being able to compare different individual designs was seen as useful.

But only about 60% of the web developers found the crowdsourced user interfaces to be useful for improving the given website to better match user needs. The remainder of the web developers were confused by the different user interfaces and found it difficult to draw actionable conclusions. These web developers focused on the shortcomings of individual crowdsourced user interfaces, as demonstrated by the comment from a web developer: *"every draft seems to adapt in a different way for certain kind of people."* Some crowdsourced user interfaces were also very similar to the original website design because only a few modifications were made. Overall, the web developers struggled to interpret the individual crowdsourced designs, and while some of the user-created modifications were deemed valuable, overall the crowdsourced user interfaces were confusing for web developers and difficult to evaluate. The individual crowdsourced user interfaces provided only limited insights.

However, a clear benefit emerged in Study 2. In this study, the crowdsourced modifications of the user interfaces were aggregated in heatmaps (depicted in Figure 8). Given the heatmaps, web developers were able to draw actionable conclusions. The heatmaps provided the web developers with an overview of the needs of a user group and an opportunity to learn about the user groups. Over 80% of the web developers agreed that their awareness of the users' needs increased after viewing the heatmaps. The heatmaps also allowed one to identify the important and unimportant parts of the user interface, as well as parts of the website that could be improved and adapted for different user groups.

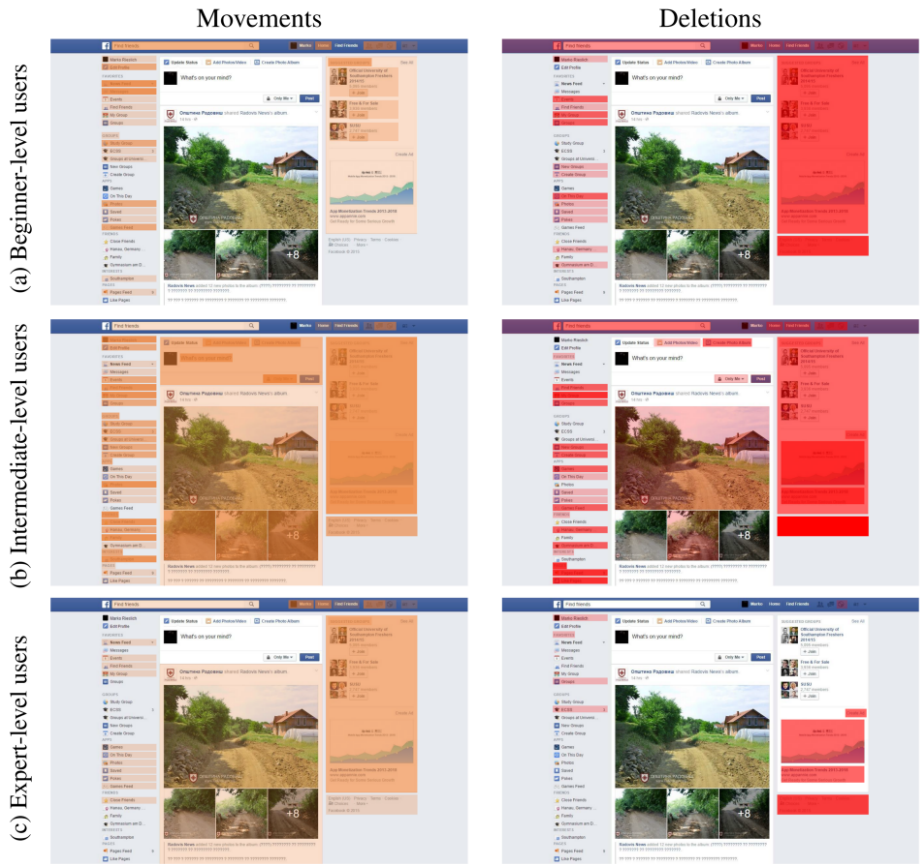

**Fig. 8. Aggregated movements (top) and deletions (bottom) of UI elements by self-rated a) beginners, b) intermediate, and c) expert users of the website. The darker the shading in the heatmap, the more users moved or deleted the respective element (reprinted with permission from Article VI © ACM, 2020).**

## 5.3 Design implications for research and practice

In this section, we provide recommendations for requesters to consider when designing creative crowdsourcing campaigns. We further give recommendations on how to adjust creative work for the different archetypes of crowd workers.

### 5.3.1 Selecting the right platform for crowdsourcing creative work

The idea of Amazon Mechanical Turk as a "market for lemons" (Akerlof, 1970; Ipeirotis, 2010c) has found some traction in academia (Ahler et al., 2019; Lykourentzou et al., 2018; Matherly, 2019). The argument is intuitive: underpayment drives high-performing crowd workers out of the marketplace, leaving only low-performing crowd workers ("lemons"). The latter group of crowd workers performs the bulk of the work on the crowdsourcing platform and, therefore, become exposed to many different types of tasks. Over-surveying may, however, also be the researcher's fault. Robinson et al. (2019) stated that "concerns about non-naivete on MTurk are due less to the MTurk platform itself and more to the way researchers use the platform." It is thus crucial for requesters of creative work to consider the right choice of platform when planning a crowdsourcing campaign.

Our research found that casual workers on Prolific were more comfortable with creative work, compared to professional crowd workers. Our findings indicate that casual workers have been less exposed to creative work and were more engaged and interested in this type of work. Prolific, therefore, may be a better pool for recruiting participants for creativity-oriented research, compared to Amazon Mechanical Turk.

In the same vein, setting high qualification criteria, as is typical in the crowdsourcing community, may be counterproductive. High qualification criteria limit the pool of participants to professional crowd workers who on the one hand have been exposed to many creativity studies (especially creativity tests) and, on the other hand, may be opposed to creative work. As recommended by Robinson et al. (2019), requesters should consider this trade-off and adapt their sampling strategies in light of this finding.

### 5.3.2 Designing tasks for creative work

Requesters should provide clear instructions for the given task to allow especially professional and pragmatic crowd workers to estimate the effort needed to complete the task. Pragmatic workers and self-developers should be informed about the goals of the study and how the collected data will be used. Self-developers and novelty seekers will naturally respond well to tasks that were designed with mini-c creativity in mind. In any case, user interfaces should be easy to use and not confusing.

While pragmatic and professional workers do not mind monotonous work, novelty seekers avoid monotony and appreciate the variety of different tasks. As a potential solution, requesters could, for instance, interleave different types of tasks (Rao, Kaur, & Lasecki, 2018) or rotate jobs (Salehi & Bernstein, 2018).

Regarding creativity tests, we suggest, instead of mandating workers to take a given creativity test, requesters could let crowd workers choose a creativity test that the worker has never encountered before. A similar strategy may be beneficial when it comes to computational priming, as detailed in Article II. The strategy of self-selection, however, comes with a trade-off since professional workers may want to seek out known tasks since such tasks can be completed more quickly with less uncertainty in the outcome.

### 5.3.3 Rewarding crowd workers for creative work

In Article I, we found a general trend of crowd workers complaining about creative work being underpaid. For compensating crowd workers fairly, requesters of creative work should be transparent about the time and effort needed to complete a task. Fair compensation, however, is only one side of the story. While general guidelines, such as paying at least a minimum wage (Silberman et al., 2018), are important for designing fair crowdsourcing campaigns, rewards that cater to the intrinsic motivation of creative crowd workers may complement traditional methods and provide intrinsic value to crowd workers. Different archetypes of crowd workers may appreciate different types of rewards. Self-developers, for instance, may appreciate receiving personal insights and may seek out tasks that help them learn new skills and knowledge. Casual workers, in general, were less concerned with monetary rewards and more interested in passing time in a fun and enjoyable manner. Requesters should therefore carefully consider how they can motivate crowd workers beyond monetary rewards. Self-developers, for instance, will greatly appreciate receiving debriefing and follow-up results, and pragmatic workers will appreciate knowing what their contributions are being used for.

### 5.3.4    Designing collaborative workflows for creative work

Collaborative ideation is the predominant paradigm in creativity-directed research in HCI today Frich et al. (2019, 2018). The preference for collaborative work could potentially affect the outcome of collaborative ideation experiments. For example, it is common for workers to self-select the tasks they work on. Self-selection bias could result from workers not wanting to participate in collaborative creative studies.

Our work found that crowd workers strongly prefer to work alone. One major pain point for crowd workers was the long latency that crowd workers anticipated in collaborative creative work, such as long response times by their counter-parts, or other technical problems that may delay the completion of the task. In the future, this area of research can be expected to benefit from technological advances, in particular the reduction of latency times. However, the underlying problem is of a socio-technical nature. The strong preference for working alone is a concern for the research area of collaborative creativity. Requesters should avoid time-traps and delays caused due to the collaborative nature of work with other crowd workers, especially if the participants include professional workers.

### 5.3.5    Aggregating complex creative work

Last, our investigation demonstrated that requesters need to be supported with inter-preting crowdsourced creative work. The need for aggregating creative work became apparent in articles IV and VI. Web design, as an example of complex creative work, can confuse both crowd workers and requesters, and requesters may have difficulties in making sense of the crowdsourced creative work.

For creative feedback, a number of different crowd feedback systems already address this issue with different visualisations. Voyant by Xu et al. (2014), for instance, aggre-gates feedback in word clouds, click clouds, and histograms. Decipher is also a system that supports requesters in interpreting crowdsourced feedback with a visualisation. But research on mechanisms for aggregating more complex creative work and creative feedback is still emerging in the feedback research community. For instance, how can we aggregate sketches, screenshots, or web designs? Our study in Article VI confirmed heat maps as a valid way of aggregating the latter. Article III found that students, in particular, valued diversity in the responses and appreciated the direct contrast between the feedback from the two sources. More research is needed on how requesters can be supported in evaluating crowdsourced creative work, and tools need to be created to better support complex creative work in general.

# 6     Discussion

This thesis explored creative work from the perspective of two key stakeholders (crowd workers and requesters) on two of the leading crowdsourcing platforms (Amazon Mechanical Turk and Prolific). The two crowdsourcing platforms have seen vast adoption in academia and industry. Anderson et al. (2019) referred to this phenomenon as the "mturkification" of research. Work is conducted on these crowdsourcing platforms in a specific ecosystem with a specific culture (Robinson, Rosenzweig, & Litman, 2020). Creative work is typically not the norm in the ecosystem of these two crowdsourcing platforms, because the platforms were primarily created for microtasks and survey research, respectively. Yet, creative work – in one form or another – is launched on these crowdsourcing platforms. This thesis provided an exploration into the perception and felt experience of creative work on such crowdsourcing platforms, and investigated how the requester can be supported in interpreting the creative work.

In this chapter, we revisit the research aims and objectives defined at the start of the thesis, and we briefly address how the thesis answered each research question. Subsequently, we outline our vision of creative work on general-purpose crowdsourcing platforms. We conclude the chapter by listing the key limitations of the thesis.

## 6.1     Exploring stakeholder perspectives of creative work

The research in this thesis is built on the human-centred auspices that the stakeholders' perspectives matter. The research aimed to explore the perspectives and experiences of creative work with a focus on two key stakeholder groups on crowdsourcing platforms: crowd workers and requesters. Our research objectives were two-fold. First, we aimed to further our understanding of the key issues that may limit creative work on general-purpose crowdsourcing platforms from the perspective of the crowd worker. Second, we created artefacts to elicit authentic experiences from both crowd workers and requesters of crowdsourced creative work.

## 6.2    Revisiting the research questions

*RQ1 – How do crowd workers perceive creative work on the crowdsourcing platform?*

Our work identified different archetypes of crowd workers. The archetypes may have an entirely different perception of and attitude towards creative work. Based on the crowd workers' demographics and motivational disposition, some crowd workers were more open to creative work than others. Casual workers enjoy and Novelty Seekers even seek out creative tasks, and Casual Workers were less exposed to common types of creative work. In the mind of professional crowd workers, creative work may lead to unforeseeable socio-technical issues and uncertain rewards. Due to this precarious nature of creative work, professional crowd workers prefer to avoid creative work and also collaborative creative work. Self-developers enjoy creative work because this type of creative work allows the worker to learn new skills or gain new knowledge about themselves. Pragmatic workers do not care about creative work and will accept to complete creative work as long as the work is well described and fairly remunerated.

Article VI found that complex creative work may confuse and overwhelm crowd workers. Creative work may, as demonstrated in Article IV, also have surprising properties that can further confuse crowd workers, if the task is not well designed. On the other hand, priming crowd workers with a potentially surprising role in Article II did not produce the expected results. Ideas from primed crowd workers were not significantly more creative than ideas from non-primed crowd workers. Our empirical work culminated in a description of the archetypal profiles of crowd workers, and a list of factors that may cause crowd workers to avoid creative work.

*RQ2 – How do requesters perceive and experience creative work?*

Our investigation of the second research question primarily focused on the end stages of the requester's creative process, which explains our focus on creative feedback as one key way of supporting requesters with crowdsourced creative work.

Article III contributed a detailed qualitative understanding of how students, as a proxy for the requester of creative feedback, perceive creative feedback. In a design-oriented University-level course, we explored the perception and felt experience of creative design feedback by students. Students appreciated the crowdsourced design feedback for being less harsh, but preferred peer design feedback over crowdsourced design feedback in all other regards. For instance, crowdsourced design feedback was found to be less specific, relevant, actionable, and qualitatively of lower value than peer design feedback. We investigated how, in the voice of the student, the quality and fairness of the creative feedback were perceived. Last, Article III contributed an investigation of the monetary valuation of crowdsourced design feedback. The analysis of the students' responses revealed a gap in how creative feedback is perceived and valued. Some of the students had a bona fide trust in the crowd workers' responses, which was driven by the students' mental image of the crowd worker being a professional worker. This belief prompted some students to attribute value to the crowdsourced feedback, even if the feedback was not particularly valuable or relevant to the student.

Article VI found that complex crowdsourced creative work may hold value to the requester (in this case, skilled web developers). However, requesters had difficulty making sense of the crowdsourced individual creative work. This problem is addressed in the last research question.

### *RQ3 – How can we support requesters in evaluating crowdsourced creative work?*

We explored this research question in three case studies in which we created computational artefacts and interfaces to support the requester in interpreting and exploring crowdsourced creative work.

In Article IV, we devised and implemented a system to elicit feedback on public displays. The preliminary evaluation with requesters revealed that complex types of creative feedback hold rich information, but do not scale well to large quantities of feedback. For instance, requesters reported that they appreciate seeing video commentaries as feedback, but would not want to evaluate a large number of such feedback items.

In Article V, we explored a faceted-filtering interface to support the exploration of ideas in a crowdsourced knowledge base. The system was found to be a valid way for finding and selecting ideas for a writing piece. Participants even preferred this solution to a well-known search engine.

In Article VI, requesters appreciated seeing the crowdsourced complex creative work (in this case, web designs). Requesters did, however, acknowledge that the individual creative works were difficult to interpret. Aggregating the complex creative work into heatmaps proved to be a viable method of supporting the requester in evaluating and interpreting the complex creative work in this case.

## 6.3    The ethical imperative for crowdsourced creative work

This research was motivated by a growing concern over the future of work, specifically job loss and job displacement associated with rapid technological advances in automation and artificial intelligence (AI). Human-labelled data is often a foundation of today's AI-based systems (Amershi, Cakmak, Knox, & Kulesza, 2014; Demartini, Difallah, Gadiraju, & Catasta, 2016). Workers on crowdsourcing platforms contribute to labelling data sets for supervised machine learning (ML) which may help machines understand, for instance, the contents of images (Simons, Gurari, & Fleischmann, 2020; von Ahn & Dabbish, 2004). However, as ML-based systems mature, the same crowd workers who work on labelling data also contribute to training machine learning models that may later make the crowd workers' jobs obsolete. The narrow focus of academia on microtask crowdsourcing and piecework (Vakharia & Lease, 2015) fails to capture the big picture in this regard. It is imperative to design systems that promote crowd workers from their current status as "ghosts" in a machinery (Gray & Suri, 2019) towards being first-class citizens in complex systems. This requires consideration of what and how humans can contribute to future value chains, and how work on general-purpose paid crowdsourcing platforms can be reorganised and orchestrated to better utilise and leverage workers' unique human abilities.

One such uniquely human ability is creativity. In studies about the future of work, creativity is listed as one advanced cognitive skill that will be in strong demand in the future (Florida, 2012; Kittur et al., 2013; Organisation for Economic Co-operation and Development (OECD), 2017; World Economic Forum, 2016). Yet, designing technologies to empower people to contribute to creative work is hard. Creativity support has been considered a grand challenge in HCI (Shneiderman, 2009; Shneiderman et al., 2006), and complex creative work, such as web design, remains "stubbornly difficult to achieve with crowdsourcing workflows" (Retelny, Bernstein, & Valentine, 2017). Complex creative work may even "remain a fundamental limitation of workflow-based crowdsourcing infrastructures" (Retelny et al., 2017). To this end, this thesis contributed towards an understanding of the design space for creative work on general-purpose paid crowdsourcing platforms: an investigation into the perspectives of the crowd worker and the requester. In the following section, we outline our vision of creative work on online paid crowdsourcing platforms.

## 6.4 Vision of creative work on crowdsourcing platforms

How can we tap into the cognitive creative surplus and ability of crowd workers without falling into the many pitfalls discussed in this thesis? In this section, we discuss our vision of creative work on general-purpose paid crowdsourcing platforms.

### Towards empowering crowd workers

Intellectual property rights are one way to empower crowd workers. Current crowdsourcing systems are very clear in this regard: crowd workers have to relinquish all intellectual property rights and idea ownership is signed over to the requester who holds the full intellectual property. Amazon Mechanical Turk (2020), for instance, states in its Participation Agreement:

> "Any work product from Tasks you perform as a Worker is a "work made for hire" for the benefit of the Requester, and you (i) agree that all ownership rights, including all intellectual property rights, will vest with that Requester immediately upon your performance of those Tasks, and (ii) waive all moral or other proprietary rights that you may have in that work product."

As outlined in Section 2.5, this may not be an issue on crowdsourcing platforms in which work predominantly is either not creative or conducted on a small-c level of creativity. Big-C creativity may be rare in such environments, but inventions with a large magnitude could nevertheless occur. Only the operators of crowdsourcing platforms hold the power to make a change in this regard. The predominant focus in the research community and industry on "human computation" and "humans as a service" (Irani & Silberman, 2013) is perhaps not going to change unless another crowdsourcing platform is created that innovates over the current standard. We describe our ideas for such a new crowdsourcing platform in the following section.

### Towards an ecosystem and a tailored platform for creative micro work

Unlocking the full potential of creative (micro) work on crowdsourcing platforms, such as Amazon Mechanical Turk, would require extensive policy changes and restructuring of workflows. Perhaps a better way of approaching the problem is by rethinking what an ecosystem for creative microwork and an entirely new tailored platform for creative microtasks could look like.

This new tailored platform should assign tasks to crowd workers based on the crowd workers' specific work preferences and cognitive skills (Hettiachchi, van Berkel, Kostakos, & Goncalves, 2020). The anonymous crowdsourcing model is ill-equipped for this task. On the other hand, even if requesters had knowledge of a worker's background, detailed knowledge of a worker's prior work and skill set would be required to accurately assign tasks to the right crowd workers.

The tailored crowdsourcing platform would therefore need to be much more transparent than current crowdsourcing platforms. Requesters need to understand a worker's work history to model the worker's preferences and optimally allocate creative work to the crowd worker. Our research further found that transparent worker profiles may be a means to alleviate misunderstandings on the side of the requester. Requesters today not only lack information about a worker's skill set, but we also lack a way of describing this skill set in general. As a potential solution, a skills taxonomy could be created to better allow to describe a crowd worker's creative work strengths, weaknesses, and work profile in general.

The tailored platform would require mechanisms to cultivate a pool of participants willing and interested in creative work. The platform could provide feedback on the quality of creative work which would be especially appreciated by self-developers. Such feedback could also, gradually over time, alleviate the crowd workers' concerns and reservations about creative work.

The platform would need to provide participants with rewards that cater to the participants' intrinsic motivation and archetypal profile. This, as detailed in Section 5.3, would require the platform to know the archetype of the crowd worker. The platform would therefore need to be able to detect the archetype of a crowd worker, for instance by classifying the crowd worker's past behaviour.

Perhaps one promising technology for a tailored platform would be blockchain technology. The crowd workers' creative contributions could be encoded in this blockchain and would, therefore, be visible and publicly accessible. Blockchain technology would allow tracking the provenance of each contribution, no matter how small. Mechanisms for sharing rewards could be devised, based on a crowd worker's creative contributions to the end product. This solution could also be a promising basis for empowering crowd workers with a mechanism to publish their reputation and experience. This would certify the crowd workers' experience, and requesters could funnel specific creative work to crowd workers who are interested in such work.

## 6.5 Limitations and reflections

In this section, we provide a brief reflection on the limitations of the thesis and the thesis author reflects on the evolution of his qualitative approach.

### 6.5.1 Limitations of the thesis

This section acknowledges a number of key limitations of the thesis.

Crowdsourcing, in general, is subject to limitations and biases. For instance, since work on crowdsourcing platforms is conducted anonymously by a diverse group of crowd workers who self-select to work on tasks, crowdsourcing studies may have low reproducibility (Paritosh, 2012) and low reliability (S.-W. Huang & Fu, 2013). Further, the crowd on crowdsourcing platforms is often not representative. Results obtained from one sample of crowd workers may not generalise to other participant samples or to other crowdsourcing platforms (Mullinix, Leeper, Druckman, & Freese, 2015). The external validity of crowdsourcing studies may, therefore, be limited. This thesis only covered two general-purpose paid crowdsourcing platforms (Amazon Mechanical Turk and Prolific). Although the two platforms are among the most popular ones, the results may not generalise to other crowdsourcing platforms.

For an unbiased analysis of the two crowdsourcing platforms, we used common qualification criteria widely used in industry and academia (Peer, Vosgerau, & Acquisti, 2014). We decided not to use post-hoc filtering and other quality control measures found in, for example, Daniel et al. (2018). We believe that introducing additional quality control mechanisms would have biased our comparison of the two crowdsourcing platforms.

The thesis follows a mixed-methods approach, and the methods used in the thesis have their own limitations. For instance, laboratory studies may have limited ecological validity and the participant's awareness of being studied may bias the results (Mc-Cambridge, Witton, & Elbourne, 2014). Articles I and III relied on empirical online surveys as the sole source of information. The responses to the questionnaires need to be considered in light of issues that are generally present in survey-based research. For instance, the collection of subjective judgements is subject to biases (Couper, 2000;

Hube, Fetahu, & Gadiraju, 2019). Crowd workers may be motivated to maximise their revenue by completing tasks as quickly as possible (Gadiraju et al., 2014; Hamby & Taylor, 2016) which may affect the data quality and its validity. As mentioned in Article I, there is also a possibility that crowd workers may have introduced bias on purpose in the hope of shifting public opinion and improving their work conditions on the crowdsourcing platform. Participants in our studies were extrinsically rewarded (with an exception in Article VI), and the participants' motivation may differ from participants who are intrinsically motivated. The latter could potentially have produced different results (D. Chandler & Kapelner, 2013).

In the second part of the thesis, research was conducted in several distinct contexts. The findings from these studies are ecologically valid in their respective context, with limited generalisability to other contexts. In Article III, our choice of study design was primarily shaped by pedagogical considerations in the context of teaching an HCI design course. In the other case studies (articles IV – VI), the context was mandated by the purpose of the study.

### 6.5.2    Reflection on the qualitative approach

When it comes to qualitative work, the thesis author's epistemological stance and approach to conducting and reporting qualitative research have evolved during this thesis. The early work in Article I and Article VI followed a "codebook" (Braun, Clarke, Hayfield, & Terry, 2019) and "small q" approach (Kidder & Fine, 1987). According to this approach, 'codes' were seen as "comments linked to extracts of text, indicating material identified by the analyst as relevant to their research question" (King, Brooks, & Tabari, 2018). The qualitative work was based on in-vivo codes (key phrases in the participants own words). Themes were constructed inductively and iteratively from the bottom-up by merging and clustering the codes into meaningful groups. The early approach to reporting this qualitative analysis was influenced by the thesis author's philosophical assumptions and positivist viewpoint. Article I, for instance, quantifies the absolute and relative frequencies of responses for each theme identified. In later work (articles III and V), the thesis author became more comfortable with qualitative work and more accepting of the fact that qualitative findings are subjectively interpreted by the researcher. The thesis author aimed to report the "big picture," and it was reported whether findings were derived from single instances or from the general trend among participants.

# 7    Conclusion

This work studied creative work on general-purpose paid crowdsourcing platforms with the broad objective of exploring the experiences and perspectives of creative work by two key stakeholders: crowd workers and requesters. The work contributes empirical insights into the perception of crowdsourced creative work and provides design implications based on different archetypal profiles of crowd workers. The research provides fruitful advice for researchers and industry professionals who wish to harness the inherent convenience and power of general-purpose paid crowdsourcing platforms for creative work (e.g. in scientific experiments, crowd feedback systems, or crowd-powered creativity support tools). However, the research is but a first step to inform the value-sensitive design of new tailored crowdsourcing platforms for creative microwork and new types of crowd feedback systems and creativity support tools that support requesters with complex creative work.

# References

Abbas, T., Khan, V.-J., Tetteroo, D., & Markopoulos, P. (2018). How creative is the crowd in describing smart home scenarios? In *Extended Abstracts of the 2018 CHI Conference on Human Factors in Computing Systems* (pp. 1–6). New York, NY, USA: Association for Computing Machinery. doi: 10.1145/3170427.3188509

Abdullah, S., Czerwinski, M., Mark, G., & Johns, P. (2016). Shining (blue) light on creative ability. In *Proceedings of the 2016 ACM International Joint Conference on Pervasive and Ubiquitous Computing* (p. 793–804). New York, NY, USA: Association for Computing Machinery. doi: 10.1145/2971648.2971751

Ahler, D. J., Roush, C. E., & Sood, G. (2019). *The micro-task market for "lemons": Data quality on Amazon's Mechanical Turk.* (Pre-print)

Akerlof, G. A. (1970). The market for "lemons": Quality uncertainty and the market mechanism. *The Quarterly Journal of Economics*, *84*(3), 488–500.

Ali, R., Hosseini, M., Phalp, K. T., & Taylor, J. (2014). The four pillars of crowdsourcing: A reference model. In *The IEEE Eighth International Conference on Research Challenges in Information Science* (p. 1-12). doi: 10.1109/RCIS.2014.6861072

Alkhatib, A., Bernstein, M. S., & Levi, M. (2017). Examining crowd work and gig work through the historical lens of piecework. In *Proceedings of the 2017 CHI Conference on Human Factors in Computing Systems* (pp. 4599–4616). New York, NY, USA: Association for Computing Machinery. doi: 10.1145/3025453.3025974

Amabile, T. M. (1983a). *The social psychology of creativity*. New York, NY, USA: Springer.

Amabile, T. M. (1983b). The social psychology of creativity: A componential conceptualization. *Journal of Personality and Social Psychology*, *45*(2), 357–376. doi: 10.1037/0022-3514.45.2.357

Amabile, T. M., Goldfarb, P., & Brackfleld, S. C. (1990). Social influences on creativity: Evaluation, coaction, and surveillance. *Creativity Research Journal*, *3*(1), 6–21. doi: 10.1080/10400419009534330

Amazon Mechanical Turk. (2020, March). *Participation agreement.* Retrieved from https://www.mturk.com/participation-agreement

Amershi, S., Cakmak, M., Knox, W. B., & Kulesza, T. (2014). Power to the people: The role of humans in interactive machine learning. *AI Magazine*, *35*(4), 105-120.

doi: 10.1609/aimag.v35i4.2513

Anderson, C. A., Allen, J. J., Plante, C., Quigley-McBride, A., Lovett, A., & Rokkum, J. N. (2019). The mturkification of social and personality psychology. *Personality and Social Psychology Bulletin*, *45*(6), 842–850. doi: 10.1177/0146167218798821

Andolina, S., Klouche, K., Cabral, D., Ruotsalo, T., & Jacucci, G. (2015). InspirationWall: Supporting idea generation through automatic information exploration. In *Proceedings of the 2015 ACM SIGCHI Conference on Creativity and Cognition* (pp. 103–106). New York, NY, USA: Association for Computing Machinery. doi: 10.1145/2757226.2757252

Andolina, S., Schneider, H., Chan, J., Klouche, K., Jacucci, G., & Dow, S. (2017). Crowdboard: Augmenting in-person idea generation with real-time crowds. In *Proceedings of the 2017 ACM SIGCHI Conference on Creativity and Cognition* (pp. 106–118). New York, NY, USA: Association for Computing Machinery. doi: 10.1145/3059454.3059477

Ashktorab, Z., Liao, Q. V., Dugan, C., Johnson, J., Pan, Q., Zhang, W., . . . Campbell, M. (2020). Human-AI collaboration in a cooperative game setting: Measuring social perception and outcomes. *Proceedings of the ACM on Human-Computer Interaction*, *4*(CSCW2). doi: 10.1145/3415167

August, T., & Reinecke, K. (2019). Pay attention, please: Formal language improves attention in volunteer and paid online experiments. In *Proceedings of the 2019 CHI Conference on Human Factors in Computing Systems* (pp. 248:1– 248:11). New York, NY, USA: Association for Computing Machinery. doi: 10.1145/3290605.3300478

Baldwin, C., & von Hippel, E. (2011). Modeling a paradigm shift: From producer innovation to user and open collaborative innovation. *Organization Science*, *22*(6), 1399–1417. doi: 10.1287/orsc.1100.0618

Behrend, T. S., Sharek, D. J., Meade, A. W., & Wiebe, E. N. (2011). The viability of crowdsourcing for survey research. *Behavior Research Methods*, *43*(3), 800–813. doi: 10.3758/s13428-011-0081-0

Bentley, F., O'Neill, K., Quehl, K., & Lottridge, D. (2020). Exploring the quality, efficiency, and representative nature of responses across multiple survey panels. In *Proceedings of the 2020 CHI Conference on Human Factors in Computing Systems* (p. 1–12). New York, NY, USA: Association for Computing Machinery. doi: 10.1145/3313831.3376671

Bernstein, M. S. (2012). *Crowd-powered systems* (Doctoral dissertation, Massachusetts Institute of Technology). DSpace@MIT. Retrieved from `http://hdl.handle.net/1721.1/74888`

Bernstein, M. S., Little, G., Miller, R. C., Hartmann, B., Ackerman, M. S., Karger, D. R., ... Panovich, K. (2010). Soylent: A word processor with a crowd inside. In *Proceedings of the 23rd Annual ACM Symposium on User Interface Software and Technology* (pp. 313–322). New York, NY, USA: Association for Computing Machinery. doi: 10.1145/1866029.1866078

Bernstein, M. S., Teevan, J., Dumais, S., Liebling, D., & Horvitz, E. (2012). Direct answers for search queries in the long tail. In *Proceedings of the SIGCHI Conference on Human Factors in Computing Systems* (p. 237–246). New York, NY, USA: Association for Computing Machinery. doi: 10.1145/2207676.2207710

Boden, M. A. (1991). *The creative mind: Myths and mechanisms.* New York, NY, USA: Basic Books, Inc.

Boden, M. A. (1996). Chapter 9 – creativity. In M. A. Boden (Ed.), *Artificial intelligence* (pp. 267–291). San Diego: Academic Press. doi: 10.1016/B978-012161964-0/50011-X

Bohannon, J. (2011). Social science for pennies. *Science*, *334*(6054), 307–307. doi: 10.1126/science.334.6054.307

Borning, A., & Muller, M. (2012). Next steps for value sensitive design. In *Proceedings of the SIGCHI Conference on Human Factors in Computing Systems* (p. 1125–1134). New York, NY, USA: Association for Computing Machinery. doi: 10.1145/2207676.2208560

Braun, V., & Clarke, V. (2006). Using thematic analysis in psychology. *Qualitative Research in Psychology*, *3*(2), 77–101. doi: 10.1191/1478088706qp063oa

Braun, V., Clarke, V., Hayfield, N., & Terry, G. (2019). Thematic analysis. In P. Liamputtong (Ed.), *Handbook of research methods in health social sciences* (p. 843–860). Singapore: Springer. doi: 10.1007/978-981-10-5251-4_103

Brown, T. B., Mann, B., Ryder, N., Subbiah, M., Kaplan, J., Dhariwal, P., ... Amodei, D. (2020). *Language models are few-shot learners.* arXiv pre-print 2005.14165. Retrieved from `https://arxiv.org/abs/2005.14165`

Buhrmester, M., Kwang, T., & Gosling, S. D. (2011). Amazon's Mechanical Turk: A new source of inexpensive, yet high-quality, data? *Perspectives on Psychological Science*, *6*(1), 3-5. (PMID: 26162106) doi: 10.1177/1745691610393980

Buisine, S., Guegan, J., Barré, J., Segonds, F., & Aoussat, A. (2016). Using avatars to tailor ideation process to innovation strategy. *Cognition, Technology & Work*, *18*(3), 583–594. doi: 10.1007/s10111-016-0378-y

Cai, C. J., Jongejan, J., & Holbrook, J. (2019). The effects of example-based explanations in a machine learning interface. In *Proceedings of the 24th International Conference on Intelligent User Interfaces* (pp. 258–262). New York, NY, USA: Association for Computing Machinery. doi: 10.1145/3301275.3302289

Chan, J., Siangliulue, P., Qori McDonald, D., Liu, R., Moradinezhad, R., Aman, S., . . . Dow, S. P. (2017). Semantically far inspirations considered harmful?: Accounting for cognitive states in collaborative ideation. In *Proceedings of the 2017 ACM SIGCHI Conference on Creativity and Cognition* (pp. 93–105). New York, NY, USA: Association for Computing Machinery. doi: 10.1145/3059454.3059455

Chandler, D., & Kapelner, A. (2013). Breaking monotony with meaning: Motivation in crowdsourcing markets. *Journal of Economic Behavior & Organization*, *90*, 123–133. doi: 10.1016/j.jebo.2013.03.003

Chandler, J., Mueller, P., & Paolacci, G. (2014). Nonnaïveté among Amazon Mechanical Turk workers: Consequences and solutions for behavioral researchers. *Behavior Research Methods*, *46*(1), 112–130. doi: 10.3758/s13428-013-0365-7

Chandler, J., Paolacci, G., Peer, E., Mueller, P., & Ratliff, K. A. (2015). Using nonnaive participants can reduce effect sizes. *Psychological Science*, *26*(7), 1131–1139. doi: 10.1177/0956797615585115

Chandler, J., & Shapiro, D. (2016). Conducting clinical research using crowdsourced convenience samples. *Annual Review of Clinical Psychology*, *12*(1), 53–81. doi: 10.1146/annurev-clinpsy-021815-093623

Colton, S., & Wiggins, G. A. (2012). Computational creativity: The final frontier? In L. De Raedt et al. (Eds.), *Frontiers in artificial intelligence and applications* (Vol. 242, p. 21-26). doi: 10.3233/978-1-61499-098-7-21

Couper, M. P. (2000). Web surveys: A review of issues and approaches. *Public Opinion Quarterly*, *64*(4), 464–494. doi: 10.1086/318641

Cranshaw, J., & Kittur, A. (2011). The polymath project: Lessons from a successful online collaboration in mathematics. In *Proceedings of the SIGCHI Conference on Human Factors in Computing Systems* (p. 1865–1874). New York, NY, USA: Association for Computing Machinery. doi: 10.1145/1978942.1979213

Cropley, A. J. (2011). Definitions of creativity. In M. A. Runco & R. Pritzker Steven (Eds.), *Encyclopedia of Creativity* (2nd ed., Vol. 1, pp. 358–368). Academic Press.

Daniel, F., Kucherbaev, P., Cappiello, C., Benatallah, B., & Allahbakhsh, M. (2018). Quality control in crowdsourcing: A survey of quality attributes, assessment techniques, and assurance actions. *ACM Computing Surveys*, *51*(1). doi: 10.1145/3148148

Demartini, G., Difallah, D. E., Gadiraju, U., & Catasta, M. (2016). An introduction to hybrid human-machine information systems. In *Foundations and trends in web science* (Vol. 7, p. 1–87). Hanover, MA, USA: now Publishers Inc. doi: 10.1561/1800000025

Deng, X. N., Joshi, K. D., & Galliers, R. D. (2016). The duality of empowerment and marginalization in microtask crowdsourcing: Giving voice to the less powerful through value sensitive design. *MIS Quarterly*, *40*(2), 279–302.

Dennis, A. R., Minas, R. K., & Bhagwatwar, A. P. (2013). Sparking creativity: Improving electronic brainstorming with individual cognitive priming. *Journal of Management Information Systems*, *29*(4), 195–216. doi: 10.2753/MIS0742-1222290407

Dennis, A. R., & Williams, M. L. (2003). Electronic brainstorming: Theory, research, and future directions. In P. B. Paulus & B. A. Nijstad (Eds.), *Group Creativity: Innovation through Collaboration* (pp. 160–178). New York, NY, USA: Oxford University Press. doi: 10.1093/acprof:oso/9780195147308.003.0008

Design Council. (2007). *Eleven lessons: Managing design in eleven global brands. A study of the design process.* Report. London, UK. Retrieved from https://www.designcouncil.org.uk/resources/report/11-lessons-managing-design-global-brands

Difallah, D., Catasta, M., Demartini, G., & Cudré-Mauroux, P. (2014). Scaling-up the crowd: Micro-task pricing schemes for worker retention and latency improvement. *Proceedings of the AAAI Conference on Human Computation and Crowdsourcing*, *2*(1), 50-58.

Difallah, D., Filatova, E., & Ipeirotis, P. G. (2018). Demographics and dynamics of Mechanical Turk workers. In *Proceedings of the Eleventh ACM International Conference on Web Search and Data Mining* (pp. 135–143). New York, NY, USA: Association for Computing Machinery. doi: 10.1145/3159652.3159661

Doan, A., Ramakrishnan, R., & Halevy, A. Y. (2011). Crowdsourcing systems on the World-Wide Web. *Communications of the ACM*, *54*(4), 86–96. doi: 10.1145/1924421.1924442

Dontcheva, M., Gerber, E., & Lewis, S. (2011). Crowdsourcing and creativity. In *Proceedings of the Workshop on Crowdsourcing and Human Computation: Systems, Studies and Platforms* (pp. 1–4). New York, NY, USA: Association for Computing Machinery.

Dow, S., Gerber, E., & Wong, A. (2013). A pilot study of using crowds in the classroom. In *Proceedings of the SIGCHI Conference on Human Factors in Computing Systems* (p. 227-236). New York, NY, USA: Association for Computing Machinery. doi: 10.1145/2470654.2470686

Dow, S., Kulkarni, A., Klemmer, S., & Hartmann, B. (2012). Shepherding the crowd yields better work. In *Proceedings of the ACM 2012 Conference on Computer Supported Cooperative Work* (p. 1013–1022). New York, NY, USA: Association for Computing Machinery. doi: 10.1145/2145204.2145355

Duncker, K. (1945). On problem-solving. *Psychological Monographs*, *58*(5). doi: 10.1037/h0093599

Engelbart, D. C. (1962). *Augmenting human intellect: A conceptual framework.* Summary report. Contract AF 49 638 1024 SRI Project 3578. Menlo Park, CA, USA: Stanford Research Institute.

Estellés-Arolas, E., & de Guevara, F. G.-L. (2012). Towards an integrated crowdsourcing definition. *Journal of Information Science*, *38*(2), 189-200. doi: 10.1177/0165551512437638

Faridani, S., Hartmann, B., & Ipeirotis, P. G. (2011). What's the right price? Pricing tasks for finishing on time. In *Proceedings of the 11th AAAI Conference on Human Computation* (p. 26–31). AAAI Press.

Florida, R. (2012). *The rise of the creative class – revisited.* Basic Books.

Foong, E., Dow, S. P., Bailey, B. P., & Gerber, E. M. (2017). Online feedback exchange: A framework for understanding the socio-psychological factors. In *Proceedings of the 2017 CHI Conference on Human Factors in Computing Systems* (pp. 4454–4467). New York, NY: Association for Computing Machinery. doi: 10.1145/3025453.3025791

Frey, B. S., & Jegen, R. (2001). Motivation crowding theory. *Journal of Economic Surveys*, *15*(5), 589-611. doi: 10.1111/1467-6419.00150

Frich, J., MacDonald Vermeulen, L., Remy, C., Biskjaer, M. M., & Dalsgaard, P. (2019). Mapping the landscape of creativity support tools in HCI. In *Proceedings of the 2019 CHI Conference on Human Factors in Computing Systems* (pp. 389:1–389:18). New York, NY, USA: Association for Computing Machinery. doi: 10.1145/3290605.3300619

Frich, J., Mose Biskjaer, M., & Dalsgaard, P. (2018). Twenty years of creativity research in human-computer interaction: Current state and future directions. In *Proceedings of the 2018 Designing Interactive Systems Conference* (pp. 1235–1257). New York, NY, USA: Association for Computing Machinery. doi: 10.1145/3196709.3196732

Friedman, B. (1996). Value-sensitive design. *Interactions*, *3*(6), 16–23. doi: 10.1145/242485.242493

Friedman, B., & Hendry, D. G. (2019). *Value sensitive design*. MIT Press.

Gadiraju, U., & Demartini, G. (2019). Understanding worker moods and reactions to rejection in crowdsourcing. In *Proceedings of the 30th ACM Conference on Hypertext and Social Media* (p. 211–220). New York, NY, USA: Association for Computing Machinery. doi: 10.1145/3342220.3343644

Gadiraju, U., Kawase, R., & Dietze, S. (2014). A taxonomy of microtasks on the web. In *Proceedings of the 25th ACM Conference on Hypertext and Social Media* (pp. 218–223). New York, NY, USA: Association for Computing Machinery. doi: 10.1145/2631775.2631819

Gadiraju, U., Kawase, R., Dietze, S., & Demartini, G. (2015). Understanding malicious behavior in crowdsourcing platforms: The case of online surveys. In *Proceedings of the 33rd Annual ACM Conference on Human Factors in Computing Systems* (pp. 1631–1640). New York, NY, USA: Association for Computing Machinery. doi: 10.1145/2702123.2702443

Gadiraju, U., Möller, S., Nöllenburg, M., Saupe, D., Egger-Lampl, S., Archambault, D., & Fisher, B. (2017). Crowdsourcing versus the laboratory: Towards human-centered experiments using the crowd. In D. Archambault, H. Purchase, & T. Hoßfeld (Eds.), *Evaluation in the crowd. crowdsourcing and human-centered experiments* (pp. 6–26). Cham, Switzerland: Springer International Publishing. doi: 10.1007/978-3-319-66435-4_2

Gao, Y., & Parameswaran, A. (2014). Finish them! pricing algorithms for human computation. *Proceedings of the VLDB Endowment*, *7*(14), 1965–1976. doi: 10.14778/2733085.2733101

Glaser, B. G., & Strauss, A. L. (1967). *The discovery of grounded theory. strategies for qualitative research*. New Brunswick, NJ, USA and London, UK: AldineTransaction.

Gleibs, I. H. (2017). Are all "research fields" equal? Rethinking practice for the use of data from crowdsourcing market places. *Behavior Research Methods*, *49*(4),

1333–1342. doi: 10.3758/s13428-016-0789-y

Gray, M. L., & Suri, S. (2019). *Ghost work: How to stop Silicon Valley from building a new global underclass*. Eamon Dolan Books.

Gray, M. L., Suri, S., Ali, S. S., & Kulkarni, D. (2016). The crowd is a collaborative network. In *Proceedings of the 19th ACM Conference on Computer-Supported Cooperative Work & Social Computing* (p. 134–147). New York, NY, USA: Association for Computing Machinery. doi: 10.1145/2818048.2819942

Guilford, J. P. (1950). Creativity. *American Psychologist*, *5*, 444–454.

Guilford, J. P. (1956). The structure of intellect. *Psychological Bulletin*, *53*(4), 267–293. doi: 10.1037/h0040755

Guilford, J. P. (1959). Three faces of intellect. *American Psychologist*, *14*(8), 469–479. doi: 10.1037/h0046827

Guilford, J. P. (1967). *The nature of human intelligence*. New York, NY, USA: McGraw-Hill.

Guilford, J. P., Christensen, P. R., Merrifield, P. R., & Wilson, R. C. (1978). *Alternate uses: Manual of instructions and interpretation*. Orange, CA, USA: Sheridan Psychological Services.

Guzdial, M. (2013). Human-centered computing: A new degree for Licklider's world. *Communications of the ACM*, *56*(5), 32–34. doi: 10.1145/2447976.2447987

Haigh, M. (2016). Has the standard cognitive reflection test become a victim of its own success? *Advances in Cognitive Psychology*, *12*(3), 145–149. doi: 10.5709/acp-0193-5

Hamby, T., & Taylor, W. (2016). Survey satisficing inflates reliability and validity measures: An experimental comparison of college and Amazon Mechanical Turk samples. *Educational and Psychological Measurement*, *76*(6), 912-932. doi: 10.1177/0013164415627349

Hara, K., Adams, A., Milland, K., Savage, S., Callison-Burch, C., & Bigham, J. P. (2018). A data-driven analysis of workers' earnings on Amazon Mechanical Turk. In *Proceedings of the 2018 CHI Conference on Human Factors in Computing Systems* (pp. 449:1–449:14). New York, NY, USA: Association for Computing Machinery. doi: 10.1145/3173574.3174023

Hettiachchi, D., van Berkel, N., Kostakos, V., & Goncalves, J. (2020, October). CrowdCog: A cognitive skill based system for heterogeneous task assignment and recommendation in crowdsourcing. *Proceedings of the ACM on Human-Computer Interaction*, *4*(CSCW2). doi: 10.1145/3415181

Hong, L., & Page, S. E. (2004). Groups of diverse problem solvers can outperform groups of high-ability problem solvers. *Proceedings of the National Academy of Sciences*, *101*(46), 16385–16389. doi: 10.1073/pnas.0403723101

Hornbæk, K., Sander, S. S., Bargas-Avila, J. A., & Grue Simonsen, J. (2014). Is once enough? On the extent and content of replications in human-computer interaction. In *Proceedings of the SIGCHI Conference on Human Factors in Computing Systems* (p. 3523–3532). New York, NY, USA: Association for Computing Machinery. doi: 10.1145/2556288.2557004

Horton, J. J., Rand, D. G., & Zeckhauser, R. J. (2011). The online laboratory: Conducting experiments in a real labor market. *Experimental Economics*, *14*(3), 399–425. doi: 10.1007/s10683-011-9273-9

Hosio, S., Goncalves, J., Anagnostopoulos, T., & Kostakos, V. (2016). Leveraging wisdom of the crowd for decision support. In *Proceedings of the 30th International BCS Human Computer Interaction Conference: Fusion!* Swindon, UK: BCS Learning & Development Ltd. doi: 10.14236/ewic/HCI2016.38

Hosseini, M., Shahri, A., Phalp, K., Taylor, J., & Ali, R. (2015). Crowdsourcing: A taxonomy and systematic mapping study. *Computer Science Review*, *17*, 43 - 69. doi: 10.1016/j.cosrev.2015.05.001

Howe, J. (2006). The rise of crowdsourcing. *Wired Magazine*, *14*(6), 1–4.

Huang, C.-Y., Huang, S.-H., & Huang, T.-H. K. (2020). Heteroglossia: In-situ story ideation with the crowd. In *Proceedings of the 2020 CHI Conference on Human Factors in Computing Systems* (p. 1–12). New York, NY, USA: Association for Computing Machinery. doi: 10.1145/3313831.3376715

Huang, S.-W., & Fu, W.-T. (2013). Enhancing reliability using peer consistency evaluation in human computation. In *Proceedings of the 2013 Conference on Computer Supported Cooperative Work* (pp. 639–648). New York, NY, USA: Association for Computing Machinery. doi: 10.1145/2441776.2441847

Hube, C., Fetahu, B., & Gadiraju, U. (2019). Understanding and mitigating worker biases in the crowdsourced collection of subjective judgments. In *Proceedings of the 2019 CHI Conference on Human Factors in Computing Systems* (p. 1–12). New York, NY, USA: Association for Computing Machinery. doi: 10.1145/3290605.3300637

Hui, J., Glenn, A., Jue, R., Gerber, E., & Dow, S. (2015). Using anonymity and communal efforts to improve quality of crowdsourced feedback. In *Third AAAI Conference on Human Computation and Crowdsourcing.* AAAI.

Ipeirotis, P. G. (2010a). Analyzing the Amazon Mechanical Turk marketplace. *XRDS*, *17*(2), 16–21. doi: 10.1145/1869086.1869094

Ipeirotis, P. G. (2010b). *Demographics of Mechanical Turk* (Tech. Rep. No. CeDER-10-01). New York University. Retrieved from `https://ssrn.com/abstract=1585030`

Ipeirotis, P. G. (2010c, July 27). *Mechanical Turk, low wages, and the market for lemons.* Blog post. Retrieved from `https://www.behind-the-enemy-lines.com/2010/07/mechanical-turk-low-wages-and-market.html`

Irani, L. C., & Silberman, M. S. (2013). Turkopticon: Interrupting worker invisibility in Amazon Mechanical Turk. In *Proceedings of the SIGCHI Conference on Human Factors in Computing Systems* (pp. 611–620). New York, NY, USA: Association for Computing Machinery. doi: 10.1145/2470654.2470742

Kang, H. B., Amoako, G., Sengupta, N., & Dow, S. P. (2018). Paragon: An online gallery for enhancing design feedback with visual examples. In *Proceedings of the 2018 CHI Conference on Human Factors in Computing Systems.* New York, NY: Association for Computing Machinery. doi: 10.1145/3173574.3174180

Karger, D. R., Oh, S., & Shah, D. (2014). Budget-optimal task allocation for reliable crowdsourcing systems. *Operations Research*, *62*(1), 1–24. doi: 10.1287/opre.2013.1235

Kaufman, J. C., & Beghetto, R. A. (2009). Beyond big and little: The four c model of creativity. *Review of General Psychology*, *13*(1), 1–12. doi: 10.1037/a0013688

Kaufman, J. C., & Sternberg, R. J. (2010). *The Cambridge handbook of creativity*. New York, NY, USA: Cambridge University Press.

Kees, J., Berry, C., Burton, S., & Sheehan, K. (2017). An analysis of data quality: Professional panels, student subject pools, and Amazon's Mechanical Turk. *Journal of Advertising*, *46*(1), 141–155. doi: 10.1080/00913367.2016.1269304

Kerne, A., Koh, E., Smith, S., Choi, H., Graeber, R., & Webb, A. (2007). Promoting emergence in information discovery by representing collections with composition. In *Proceedings of the 6th ACM SIGCHI Conference on Creativity & Cognition* (p. 117–126). New York, NY, USA: Association for Computing Machinery. doi: 10.1145/1254960.1254977

Kerne, A., & Smith, S. M. (2004). The information discovery framework. In *Proceedings of the 5th Conference on Designing Interactive Systems: Processes, Practices, Methods, and Techniques* (pp. 357–360). New York, NY, USA: Association for

Computing Machinery. doi: 10.1145/1013115.1013179

Kidder, L. H., & Fine, M. (1987). Qualitative and quantitative methods: When stories converge. *New Directions for Program Evaluation*, *1987*(35), 57–75. doi: 10.1002/ev.1459

Kim, J., Sterman, S., Cohen, A. A. B., & Bernstein, M. S. (2017). Mechanical novel: Crowdsourcing complex work through reflection and revision. In *Proceedings of the 2017 ACM Conference on Computer Supported Cooperative Work and Social Computing* (p. 233–245). New York, NY, USA: Association for Computing Machinery. doi: 10.1145/2998181.2998196

King, N., Brooks, J., & Tabari, S. (2018). Template analysis in business and management research. In M. Ciesielska & D. Jemielniak (Eds.), *Qualitative methods in organization studies* (pp. 179–206). Cham, Switzerland: Springer International Publishing. doi: 10.1007/978-3-319-65442-3_8

Kittur, A., Chi, E. H., & Suh, B. (2008). Crowdsourcing user studies with Mechanical Turk. In *Proceedings of the SIGCHI Conference on Human Factors in Computing Systems* (pp. 453–456). New York, NY, USA: Association for Computing Machinery. doi: 10.1145/1357054.1357127

Kittur, A., Nickerson, J. V., Bernstein, M., Gerber, E. M., Shaw, A., Zimmerman, J., . . . Horton, J. (2013). The future of crowd work. In *Proceedings of the 2013 Conference on Computer Supported Cooperative Work* (pp. 1301–1318). New York, NY, USA: Association for Computing Machinery. doi: 10.1145/2441776.2441923

Kittur, A., Smus, B., Khamkar, S., & Kraut, R. E. (2011). CrowdForge: Crowdsourcing complex work. In *Proceedings of the 24th Annual ACM Symposium on User Interface Software and Technology* (p. 43–52). New York, NY, USA: Association for Computing Machinery. doi: 10.1145/2047196.2047202

Koblin, A. M. (2009). The sheep market. In *Proceedings of the Seventh ACM Conference on Creativity and Cognition* (p. 451–452). New York, NY, USA: Association for Computing Machinery. doi: 10.1145/1640233.1640348

Komarov, S., Reinecke, K., & Gajos, K. Z. (2013). Crowdsourcing performance evaluations of user interfaces. In *Proceedings of the SIGCHI Conference on Human Factors in Computing Systems* (p. 207–216). New York, NY, USA: Association for Computing Machinery. doi: 10.1145/2470654.2470684

Kou, Y., & Gray, C. M. (2017). Supporting distributed critique through interpretation and sense-making in an online creative community. *Proceedings of the ACM on Human-Computer Interaction*, *1*(CSCW). doi: 10.1145/3134695

Kozbelt, A., & Durmysheva, Y. (2007). Understanding creativity judgments of invented alien creatures: The roles of invariants and other predictors. *The Journal of Creative Behavior*, *41*(4), 223-248. doi: 10.1002/j.2162-6057.2007.tb01072.x

Kucherbaev, P., Daniel, F., Tranquillini, S., & Marchese, M. (2016). Crowdsourcing processes: A survey of approaches and opportunities. *IEEE Internet Computing*, *20*(2), 50-56. doi: 10.1109/MIC.2015.96

Lakhani, K. R., Fayard, A.-L., Levina, N., & Pokrywa, S. H. (2012). *OpenIDEO.* Harvard Business School Technology & Operations Mgt. Unit Case No. 612-066. Retrieved from https://ssrn.com/abstract=2053435

Lakhani, K. R., & Lonstein, E. (2011). Innocentive.com. *Harvard Business School General Management Unit Case*(612-026).

Law, E., Yin, M., Goh, J., Chen, K., Terry, M. A., & Gajos, K. Z. (2016). Curiosity killed the cat, but makes crowdwork better. In *Proceedings of the 2016 CHI Conference on Human Factors in Computing Systems* (p. 4098–4110). New York, NY, USA: Association for Computing Machinery. doi: 10.1145/2858036.2858144

Lazar, J., Feng, J., & Hochheiser, H. (2017). *Research methods in human-computer interaction* (2nd ed.). Cambridge, MA, USA: Morgan Kaufmann.

Lemley, M. A. (2012). The myth of the sole inventor. *Michigan Law Review*, *110*(5), 709–760.

Lewis, S., Dontcheva, M., & Gerber, E. (2011). Affective computational priming and creativity. In *Proceedings of the SIGCHI Conference on Human Factors in Computing Systems* (pp. 735–744). New York, NY, USA: Association for Computing Machinery. doi: 10.1145/1978942.1979048

Luther, K., Tolentino, J.-L., Wu, W., Pavel, A., Bailey, B. P., Agrawala, M., . . . Dow, S. P. (2015). Structuring, aggregating, and evaluating crowdsourced design critique. In *Proceedings of the ACM Conference on Computer Supported Cooperative Work & Social Computing* (pp. 473–485). New York, NY: Association for Computing Machinery. doi: 10.1145/2675133.2675283

Lykourentzou, I., Ahmed, F., Papastathis, C., Sadien, I., & Papangelis, K. (2018). When crowds give you lemons: Filtering innovative ideas using a diverse-bag-of-lemons strategy. *Proceedings of the ACM on Human-Computer Interaction*, *2*(CSCW). doi: 10.1145/3274384

Lykourentzou, I., Khan, V.-J., Papangelis, K., & Markopoulos, P. (2019). Macrotask crowdsourcing: An integrated definition. In V.-J. Khan, K. Papangelis, I. Lykourentzou, & P. Markopoulos (Eds.), *Macrotask crowdsourcing: Engaging the* *crowds to address complex problems* (pp. 1–13). Cham, Switzerland: Springer International Publishing. doi: 10.1007/978-3-030-12334-5_1

Mackeprang, M., Khiat, A., & Müller-Birn, C. (2018). Concept validation during collaborative ideation and its effect on ideation outcome. In *Extended Abstracts of the 2018 CHI Conference on Human Factors in Computing Systems* (p. 1–6). New York, NY, USA: Association for Computing Machinery. doi: 10.1145/3170427.3188485

Maier, N. (1931). Reasoning in humans. II. The solution of a problem and its appearance in consciousness. *Journal of Comparative Psychology*, *12*, 181–194. doi: 10.1037/h0071361

Mann, S., & Cadman, R. (2014). Does being bored make us more creative? *Creativity Research Journal*, *26*(2), 165–173. doi: 10.1080/10400419.2014.901073

Mao, A., Kamar, E., Chen, Y., Horvitz, E., Schwamb, M. E., Lintott, C. J., & Smith, A. M. (2013). Volunteering versus work for pay: Incentives and tradeoffs in crowdsourcing. In B. Hartman & E. Horvitz (Eds.), *Proceedings of the First AAAI Conference on Human Computation and Crowdsourcing.* AAAI.

Martin, D., Hanrahan, B. V., O'Neill, J., & Gupta, N. (2014). Being a Turker. In *Proceedings of the 17th ACM Conference on Computer Supported Cooperative Work & Social Computing* (pp. 224–235). New York, NY, USA: Association for Computing Machinery. doi: 10.1145/2531602.2531663

Mason, W., & Suri, S. (2012). Conducting behavioral research on Amazon's Mechanical Turk. *Behavior Research Methods*, *44*(1), 1–23. doi: 10.3758/s13428-011-0124-6

Matherly, T. (2019). A panel for lemons? Positivity bias, reputation systems and data quality on MTurk. *European Journal of Marketing*, *53*(2), 195–223. doi: 10.1108/EJM-07-2017-0491

McCambridge, J., Witton, J., & Elbourne, D. R. (2014). Systematic review of the hawthorne effect: New concepts are needed to study research participation effects. *Journal of Clinical Epidemiology*, *67*(3), 267-277. doi: 10.1016/j.jclinepi.2013.08.015

Mednick, S. (1962). The associative basis of the creative process. *Psychological Review*, *69*, 220–232. doi: 10.1037/h0048850

Minder, P., Seuken, S., Bernstein, A., & Zollinger, M. (2012). CrowdManager – combinatorial allocation and pricing of crowdsourcing tasks with time constraints. In *Workshop on Social Computing and User Generated Content in conjunction with ACM Conference on Electronic Commerce (ACM-EC 2012)* (pp. 1–18).

Valencia, Spain.

Mitchell, W. J., Inouye, A. S., & Blumenthal, M. S. (2003). *Beyond productivity: Information, technology, innovation, and creativity*. Washington, DC, USA: National Academy Press.

Morgan, D. L., Ataie, J., Carder, P., & Hoffman, K. (2013). Introducing dyadic interviews as a method for collecting qualitative data. *Qualitative Health Research*, *23*(9), 1276–1284. doi: 10.1177/1049732313501889

Morris, R. R., Dontcheva, M., & Gerber, E. M. (2012). Priming for better performance in microtask crowdsourcing environments. *IEEE Internet Computing*, *16*(5), 13–19. doi: 10.1109/MIC.2012.68

Mullinix, K. J., Leeper, T. J., Druckman, J. N., & Freese, J. (2015). The generalizability of survey experiments. *Journal of Experimental Political Science*, *2*(2), 109–138. doi: 10.1017/XPS.2015.19

Norwegian National Committee for Research Ethics in Science and Technology. (2016). *Guidelines for research ethics in science and technology* (2nd ed.). Oslo, Norway.

Nouri, Z., Wachsmuth, H., & Engels, G. (2020). Mining crowdsourcing problems from discussion forums of workers. In *Proceedings of the 28th International Conference on Computational Linguistics* (pp. 6264–6276). International Committee on Computational Linguistics. doi: 10.18653/v1/2020.coling-main.551

Öllinger, M., Jones, G., Faber, A. H., & Knoblich, G. (2013). Cognitive mechanisms of insight: The role of heuristics and representational change in solving the eight-coin problem. *Journal of Experimental Psychology. Learning, Memory, and Cognition*, *39*(3), 931–939. doi: 10.1037/a0029194

Oppenheim, D. (2005). Supporting creative work. *International Journal of Human-Computer Studies on Creativity and Computational Support*, *63*, 370–382.

Oppenheimer, D. M., Meyvis, T., & Davidenko, N. (2009). Instructional manipulation checks: Detecting satisficing to increase statistical power. *Journal of Experimental Social Psychology*, *45*(4), 867–872. doi: 10.1016/j.jesp.2009.03.009

Oppenlaender, J. (2020). Crowd-powered creativity support systems. In *Proceedings of 12th ACM SIGCHI Symposium on Engineering Interactive Computing Systems.* New York, NY, USA: Association for Computing Machinery. doi: 10.1145/3393672.3398646

Oppenlaender, J., Mackeprang, M., Khiat, A., Vuković, M., Goncalves, J., & Hosio, S. (2019). $DC^2S^2$: Designing crowd-powered creativity support systems. In *Extended Abstracts of the 2019 CHI Conference on Human Factors in Computing*

*Systems.* New York, NY, USA: Association for Computing Machinery. doi: 10.1145/3290607.3299027

Oppenlaender, J., Shireen, N., Mackeprang, M., Erhan, H., Goncalves, J., & Hosio, S. (2019). Crowd-powered interfaces for creative design thinking. In *Proceedings of the 2019 on Creativity and Cognition* (pp. 722–729). New York, NY, USA: Association for Computing Machinery. doi: 10.1145/3325480.3326553

Organisation for Economic Co-operation and Development (OECD). (2017). *Future of work and skills.*

Paolacci, G., & Chandler, J. (2014). Inside the Turk: Understanding Mechanical Turk as a participant pool. *Current Directions in Psychological Science*, *23*(3), 184–188. doi: 10.1177/0963721414531598

Paolacci, G., Chandler, J., & Ipeirotis, P. G. (2010). Running experiments on Amazon Mechanical Turk. *Judgment and Decision Making*, *5*(5), 411–419.

Paritosh, P. (2012). Human computation must be reproducible. In *Proceedings of the 21st International Conference on World Wide Web.*

Paritosh, P., Ipeirotis, P. G., Cooper, M., & Suri, S. (2011). The computer is the new sewing machine: Benefits and perils of crowdsourcing. In *Proceedings of the 20th International Conference Companion on World Wide Web* (pp. 325–326). New York, NY, USA: Association for Computing Machinery. doi: 10.1145/1963192.1963335

Peer, E., Brandimarte, L., Samat, S., & Acquisti, A. (2017). Beyond the Turk: Alternative platforms for crowdsourcing behavioral research. *Journal of Experimental Social Psychology*, *70*, 153–163. doi: 10.1016/j.jesp.2017.01.006

Peer, E., Vosgerau, J., & Acquisti, A. (2014). Reputation as a sufficient condition for data quality on Amazon Mechanical Turk. *Behavior Research Methods*, *46*(4), 1023–1031. doi: 10.3758/s13428-013-0434-y

Piffer, D. (2012). Can creativity be measured? An attempt to clarify the notion of creativity and general directions for future research. *Thinking Skills and Creativity*, *7*(3), 258-264. doi: 10.1016/j.tsc.2012.04.009

Plattner, H., Meinel, C., & Leifer, L. (Eds.). (2011). *Design thinking. understand – improve – apply.* Berlin, Heidelberg: Springer. doi: 10.1007/978-3-642-13757-0

Plucker, J. A., Beghetto, R. A., & Dow, G. T. (2004). Why isn't creativity more important to educational psychologists? Potentials, pitfalls, and future directions in creativity research. *Educational Psychologist*, *39*(2), 83–96. doi: 10.1207/s15326985ep3902_1

Poetz, M. K., & Schreier, M. (2012). The value of crowdsourcing: Can users really compete with professionals in generating new product ideas? *Journal of Product Innovation Management*, *29*(2), 245–256. doi: 10.1111/j.1540-5885.2011.00893.x

Qiu, S., Gadiraju, U., & Bozzon, A. (2020). Just the right mood for HIT! In M. Bielikova, T. Mikkonen, & C. Pautasso (Eds.), *Web engineering* (pp. 381–396). Cham, Switzerland: Springer International Publishing. doi: 10.1007/978-3-030-50578-3_26

Rao, A., Kaur, H., & Lasecki, W. (2018). Plexiglass: Multiplexing passive and active tasks for more efficient crowdsourcing. In *Proceedings of the AAAI Conference on Human Computation and Crowdsourcing* (Vol. 6). New York, NY, USA: AAAI.

Retelny, D., Bernstein, M. S., & Valentine, M. A. (2017). No workflow can ever be enough: How crowdsourcing workflows constrain complex work. *Proceedings of the ACM on Human-Computer Interaction*, *1*(CSCW). doi: 10.1145/3134724

Retelny, D., Robaszkiewicz, S., To, A., Lasecki, W. S., Patel, J., Rahmati, N., . . . Bernstein, M. S. (2014). Expert crowdsourcing with flash teams. In *Proceedings of the 27th Annual ACM Symposium on User Interface Software and Technology* (p. 75–85). New York, NY, USA: Association for Computing Machinery. doi: 10.1145/2642918.2647409

Rhodes, M. (1961). An analysis of creativity. *The Phi Delta Kappan*, *42*(7), 305–310. Retrieved from http://www.jstor.org/stable/20342603

Richards, R. (1990). Everyday creativity, eminent creativity, and health: "afterview"; for CRJ issues on creativity and health. *Creativity Research Journal*, *3*(4), 300-326. doi: 10.1080/10400419009534363

Rietzschel, E. F., Nijstad, B. A., & Stroebe, W. (2010). The selection of creative ideas after individual idea generation: Choosing between creativity and impact. *British Journal of Psychology*, *101*(Pt 1), 47–68. doi: 10.1348/000712609X414204

Rittel, H. W. J., & Webber, M. M. (1973). Dilemmas in a general theory of planning. *Policy Sciences*, *4*(2), 155–169. doi: 10.1007/BF01405730

Robinson, J., Rosenzweig, C., & Litman, L. (2020). The Mechanical Turk ecosystem. In L. Litman & J. Robinson (Eds.), *Conducting online research on Amazon Mechanical Turk and beyond* (p. 27-47). United Kingdom: SAGE Publications, Inc.

Robinson, J., Rosenzweig, C., Moss, A. J., & Litman, L. (2019). Tapped out or barely tapped? Recommendations for how to harness the vast and largely unused potential of the Mechanical Turk participant pool. *PLOS ONE*, *14*(12), 1–29. doi: 10.1371/journal.pone.0226394

Rogstadius, J., Kostakos, V., Kittur, A., Smus, B., Laredo, J., & Vukovic, M. (2011). An assessment of intrinsic and extrinsic motivation on task performance in crowdsourcing markets. In L. A. Adamic, R. Baeza-Yates, & S. Counts (Eds.), *Proceedings of the Fifth International Conference on Weblogs and Social Media.* The AAAI Press.

Ross, J., Irani, L., Silberman, M. S., Zaldivar, A., & Tomlinson, B. (2010). Who are the crowdworkers?: Shifting demographics in Mechanical Turk. In *CHI '10 Extended Abstracts on Human Factors in Computing Systems* (pp. 2863–2872). Association for Computing Machinery. doi: 10.1145/1753846.1753873

Roy, S. B., Lykourentzou, I., Thirumuruganathan, S., Amer-Yahia, S., & Das, G. (2013). Crowds, not drones: Modeling human factors in interactive crowdsourcing. In *1st VLDB Workshop on Databases and Crowdsourcing (DBCrowd 2013)* (pp. 39–42).

Runco, M. A. (2014). *Creativity: Theories and themes: Research, development, and practice* (2nd ed.). San Diego, CA, USA: Elsevier Academic Press.

Runco, M. A., & Jaeger, G. J. (2012). The standard definition of creativity. *Creativity Research Journal*, *24*(1), 92–96. doi: 10.1080/10400419.2012.650092

Runco, M. A., & Pritzker, R., Steven (Eds.). (2011). *Encyclopedia of creativity* (2nd ed., Vol. 1). Academic Press.

Sadler, D. R. (1989). Formative assessment and the design of instructional systems. *Instructional Science*, *18*(2), 119–144. doi: 10.1007/BF00117714

Salehi, N., & Bernstein, M. S. (2018). Hive: Collective design through network rotation. *Proceedings of the ACM on Human-Computer Interaction*, *2*(CSCW). doi: 10.1145/3274420

Salehi, N., Teevan, J., Iqbal, S., & Kamar, E. (2017). Communicating context to the crowd for complex writing tasks. In *Proceedings of the 2017 ACM Conference on Computer Supported Cooperative Work and Social Computing* (p. 1890–1901). New York, NY, USA: Association for Computing Machinery. doi: 10.1145/2998181.2998332

Sawyer, R. K. (2012). *Explaining creativity: The science of human innovation* (2nd ed.). New York, NY, USA: Oxford University Press.

Sawyer, R. K., & DeZutter, S. (2009). Distributed creativity: How collective creations emerge from collaboration. *Psychology of Aesthetics, Creativity, and the Arts*,

*3*(2), 81–92. doi: 10.1037/a0013282

Sethapakdi, T., & McCann, J. (2019). Painting with CATS: Camera-aided texture synthesis. In *Proceedings of the 2019 CHI Conference on Human Factors in Computing Systems* (p. 1–9). New York, NY, USA: Association for Computing Machinery. doi: 10.1145/3290605.3300287

Shannon, C., & Weaver, W. (1949). *The mathematical theory of communication*. Urbana, IL, USA: University of Illinois Press.

Shaw, A. D., Horton, J. J., & Chen, D. L. (2011). Designing incentives for inexpert human raters. In *Proceedings of the ACM 2011 Conference on Computer Supported Cooperative Work* (p. 275–284). New York, NY, USA: Association for Computing Machinery. doi: 10.1145/1958824.1958865

Shi, Y., Wang, Y., Qi, Y., Chen, J., Xu, X., & Ma, K.-L. (2017). IdeaWall: Improving creative collaboration through combinatorial visual stimuli. In *Proceedings of the 2017 ACM Conference on Computer Supported Cooperative Work and Social Computing* (pp. 594–603). New York, NY, USA: Association for Computing Machinery. doi: 10.1145/2998181.2998208

Shireen, N., Erhan, H., Sanchez, R., Popovic, J., Woodbury, R., & Riecke, B. E. (2011). Design space exploration in parametric systems: Analyzing effects of goal specificity and method specificity on design solutions. In *Proceedings of the 8th ACM Conference on Creativity & Cognition* (p. 249–258). New York, NY, USA: Association for Computing Machinery. doi: 10.1145/2069618.2069660

Shneiderman, B. (2002). Creativity support tools. *Communications of the ACM*, *45*(10), 116–120. doi: 10.1145/570907.570945

Shneiderman, B. (2007). Creativity support tools: Accelerating discovery and innovation. *Communications of the ACM*, *50*(12), 20–32. doi: 10.1145/1323688.1323689

Shneiderman, B. (2009). Creativity support tools: A grand challenge for HCI researchers. In M. Redondo, C. Bravo, & M. Ortega (Eds.), *Engineering the User Interface: From Research to Practice* (pp. 1–9). London, UK: Springer. doi: 10.1007/978-1-84800-136-7_1

Shneiderman, B., Fischer, G., Czerwinski, M., Resnick, M., Myers, B., Candy, L., . . . Terry, M. (2006). Creativity support tools: Report from a U.S. National Science Foundation sponsored workshop. *International Journal of Human–Computer Interaction*, *20*(2), 61–77. doi: 10.1207/s15327590ijhc2002_1

Siangliulue, P., Arnold, K. C., Gajos, K. Z., & Dow, S. P. (2015). Toward collaborative ideation at scale: Leveraging ideas from others to generate more creative and diverse ideas. In *Proceedings of the 18th ACM Conference on Computer Supported Cooperative Work & Social Computing* (pp. 937–945). New York, NY, USA: Association for Computing Machinery. doi: 10.1145/2675133.2675239

Siangliulue, P., Chan, J., Gajos, K. Z., & Dow, S. P. (2015). Providing timely examples improves the quantity and quality of generated ideas. In *Proceedings of the 2015 ACM SIGCHI Conference on Creativity and Cognition* (pp. 83–92). New York, NY, USA: Association for Computing Machinery. doi: 10.1145/2757226.2757230

Silberman, M. S. (2015). *Stop citing Ross et al. 2010, "Who are the crowdworkers?"*. Comment on the ACM Digital Library, reproduced on Medium.com. Retrieved from `http://bit.ly/2FkrObs`

Silberman, M. S., Irani, L., & Ross, J. (2010). Ethics and tactics of professional crowdwork. *XRDS*, *17*(2), 39–43. doi: 10.1145/1869086.1869100

Silberman, M. S., Ross, J., Irani, L., & Tomlinson, B. (2010). Sellers' problems in human computation markets. In *Proceedings of the ACM SIGKDD Workshop on Human Computation* (p. 18–21). New York, NY, USA: Association for Computing Machinery. doi: 10.1145/1837885.1837891

Silberman, M. S., Tomlinson, B., LaPlante, R., Ross, J., Irani, L. C., & Zaldivar, A. (2018). Responsible research with crowds: Pay crowdworkers at least minimum wage. *Communications of the ACM*, *61*(3), 39–41. doi: 10.1145/3180492

Simon, H. A. (1956). Rational choice and the structure of the environment. *Psychological Review*, *63*(2), 129–138. doi: 10.1037/h0042769

Simon, H. A. (1996). *The sciences of the artificial* (3rd ed.). MIT Press.

Simons, R. N., Gurari, D., & Fleischmann, K. R. (2020, October). "i hope this is helpful": Understanding crowdworkers' challenges and motivations for an image description task. *Proceedings of the ACM on Human-Computer Interaction*, *4*(CSCW2). doi: 10.1145/3415176

Simonton, D. K. (2013). *Genius, creativity, and leadership. historiometric inquiries*. Harvard University Press. doi: 10.4159/harvard.9780674424753

Singer, Y., & Mittal, M. (2011). Pricing tasks in online labor markets. In *Proceedings of the 2011 AAAI Workshop (WS-11-11)* (p. 55–60). AAAI.

Singer, Y., & Mittal, M. (2013). Pricing mechanisms for crowdsourcing markets. In *Proceedings of the 22nd International Conference on World Wide Web* (p. 1157–1166). New York, NY, USA: Association for Computing Machinery. doi: 10.1145/2488388.2488489

Stephanidis, C., Salvendy, G., Antona, M., Chen, J. Y. C., Dong, J., Duffy,

V. G., . . . Zhou, J. (2019). Seven HCI grand challenges. *International Journal of Human–Computer Interaction*, *35*(14), 1229-1269. doi: 10.1080/10447318.2019.1619259

Sternberg, R. J. (2005). Creativity or creativities? *International Journal of Human-Computer Studies*, *63*(4), 370-382. doi: 10.1016/j.ijhcs.2005.04.003

Stewart, N., Chandler, J., & Paolacci, G. (2017). Crowdsourcing samples in cognitive science. *Trends in Cognitive Sciences*, *21*(10), 736–748. doi: 10.1016/j.tics.2017.06.007

Stewart, N., Ungemach, C., Harris, A. J. L., Bartels, D. M., Newell, B. R., Paolacci, G., & Chandler, J. (2015). The average laboratory samples a population of 7,300 Amazon Mechanical Turk workers. *Judgment and Decision Making*, *10*(5), 479–491.

Surowiecki, J. (2005). *The wisdom of crowds*. New York, NY, USA: Anchor.

Tashakkori, A., & Creswell, J. W. (2007). Editorial: Exploring the nature of research questions in mixed methods research. *Journal of Mixed Methods Research*, *1*(3), 207-211. doi: 10.1177/1558689807302814

Teevan, J., & Yu, L. (2017). Bringing the wisdom of the crowd to an individual by having the individual assume different roles. In *Proceedings of the 2017 ACM SIGCHI Conference on Creativity and Cognition* (pp. 131–135). New York, NY, USA: Association for Computing Machinery. doi: 10.1145/3059454.3059467

Torrance, E. P. (1966). *Torrance tests of creative thinking. Norms-technical manual. Research edition. Verbal tests, forms A and B. Figural tests, forms A and B.* Princeton, NJ, USA: Personnel Press.

Toxtli, C., Richmond-Fuller, A., & Savage, S. (2020). Reputation agent: Prompting fair reviews in gig markets. In *Proceedings of The Web Conference 2020* (p. 1228–1240). New York, NY, USA: Association for Computing Machinery. doi: 10.1145/3366423.3380199

Tsai, G. T. (2009). *Design for surprise and idea generation methods.* Master thesis. Massachusetts Institute of Technology.

Tsai, G. T. (2016). *The tools we use: A study of user preferences for sketches, prototypes, and CAD models and the influence on design outcome* (Doctoral dissertation, Massachusetts Institute of Technology). DSpace@MIT. Retrieved from http://hdl.handle.net/1721.1/106785

Vaish, R., Gaikwad, S. N. S., Kovacs, G., Veit, A., Krishna, R., Arrieta Ibarra, I., . . . Bernstein, M. S. (2017). Crowd research: Open and scalable university laboratories. In *Proceedings of the 30th Annual ACM Symposium on User Interface Software and Technology* (p. 829–843). New York, NY, USA: Association for Computing Machinery. doi: 10.1145/3126594.3126648

Vakharia, D., & Lease, M. (2015). Beyond Mechanical Turk: An analysis of paid crowd work platforms. In *iConference.*

Valentine, M. A., Retelny, D., To, A., Rahmati, N., Doshi, T., & Bernstein, M. S. (2017). Flash organizations: Crowdsourcing complex work by structuring crowds as organizations. In *Proceedings of the 2017 CHI Conference on Human Factors in Computing Systems* (p. 3523–3537). New York, NY, USA: Association for Computing Machinery. doi: 10.1145/3025453.3025811

Villalba Garcia, E. (2008). *On creativity towards an understanding of creativity and its measurements.* OPOCE. doi: 10.2788/2936

von Ahn, L., & Dabbish, L. (2004). Labeling images with a computer game. In *Proceedings of the SIGCHI Conference on Human Factors in Computing Systems* (p. 319–326). New York, NY, USA: Association for Computing Machinery. doi: 10.1145/985692.985733

Wallas, G. (1926). *The art of thought*. New York, NY, USA: Harcourt, Brace and Company.

Wang, K., & Nickerson, J. V. (2017). A literature review on individual creativity support systems. *Computers in Human Behavior*, *74*, 139–151. doi: 10.1016/j.chb.2017.04.035

Wang, Y., Papangelis, K., Lykourentzou, I., Liang, H.-N., Sadien, I., Demerouti, E., & Khan, V.-J. (2020). In their shoes: A structured analysis of job demands, resources, work experiences, and platform commitment of crowdworkers in China. *Proceedings of the ACM on Human-Computer Interaction*, *4*(GROUP). doi: 10.1145/3375187

Wang, Y., Papangelis, K., Saker, M., Lykourentzou, I., Chamberlain, A., & Khan, V.-J. (2020). Crowdsourcing in China: Exploring the work experiences of solo crowdworkers and crowdfarm workers. In *Proceedings of the 2020 CHI Conference on Human Factors in Computing Systems* (p. 1–13). New York, NY, USA: Association for Computing Machinery. doi: 10.1145/3313831.3376473

Ward, T. B. (1994). Structured imagination: The role of category structure in exemplar generation. *Cognitive Psychology*, *27*(1), 1–40.

Wauck, H., Yen, Y.-C. G., Fu, W.-T., Gerber, E., Dow, S. P., & Bailey, B. P. (2017). From in the class or in the wild? Peers provide better design feedback than external crowds. In *Proceedings of the 2017 CHI Conference on Human Factors in Computing Systems* (p. 5580–5591). New York, NY, USA: Association for Computing Machinery. doi: 10.1145/3025453.3025477

Weisberg, R. W. (2006). *Creativity: Understanding innovation in problem solving, science, invention and the arts*. Hoboken, NJ, USA: John Wiley & Sons.

Whiting, M. E., Gamage, D., Goyal, S., Gilbee, A., Majeti, D., Richmond-Fuller, A., . . . Bernstein, M. S. (2017). Designing a constitution for a self-governing crowdsourcing marketplace. In *Collective Intelligence Conference.*

Wolf, C. G., Carroll, J. M., Landauer, T. K., John, B. E., & Whiteside, J. (1989). The role of laboratory experiments in HCI: Help, hindrance, or ho-hum? In *Proceedings of the SIGCHI Conference on Human Factors in Computing Systems* (p. 265–268). New York, NY, USA: Association for Computing Machinery. doi: 10.1145/67449.67500

World Economic Forum. (2016). *The future of jobs. employment, skills and workforce strategy for the fourth industrial revolution.* Global Challenge Insight Report, Ref. 010116.

Xu, A., Huang, S.-W., & Bailey, B. (2014). Voyant: Generating structured feedback on visual designs using a crowd of non-experts. In *Proceedings of the 17th ACM Conference on Computer Supported Cooperative Work & Social Computing* (pp. 1433–1444). New York, NY: Association for Computing Machinery. doi: 10.1145/2531602.2531604

Yen, Y.-C. G., Kim, J. O., & Bailey, B. P. (2020). Decipher: An interactive visualization tool for interpreting unstructured design feedback from multiple providers. In *Proceedings of the 2020 CHI Conference on Human Factors in Computing Systems* (p. 1–13). New York, NY, USA: Association for Computing Machinery. doi: 10.1145/3313831.3376380

Yin, M., Gray, M. L., Suri, S., & Vaughan, J. W. (2016). The communication network within the crowd. In *Proceedings of the 25th International Conference on World Wide Web* (p. 1293–1303). Geneva, Switzerland: International World Wide Web Conferences Steering Committee. doi: 10.1145/2872427.2883036

Yu, L., Kittur, A., & Kraut, R. E. (2014). Distributed analogical idea generation: Inventing with crowds. In *Proceedings of the SIGCHI Conference on Human Factors in Computing Systems* (pp. 1245–1254). New York, NY, USA: Association for Computing Machinery. doi: 10.1145/2556288.2557371

Yu, L., Kittur, A., & Kraut, R. E. (2016). Encouraging "outside-the-box" thinking in crowd innovation through identifying domains of expertise. In *Proceedings of the 19th ACM Conference on Computer-Supported Cooperative Work & Social Computing* (pp. 1214–1222). New York, NY, USA: Association for Computing Machinery. doi: 10.1145/2818048.2820025

Yu, L., & Nickerson, J. V. (2011). Cooks or cobblers?: Crowd creativity through combination. In *Proceedings of the SIGCHI Conference on Human Factors in Computing Systems* (pp. 1393–1402). New York, NY, USA: Association for Computing Machinery. doi: 10.1145/1978942.1979147

Yu, L., & Nickerson, J. V. (2013). An internet-scale idea generation system. *ACM Transactions on Interactive Intelligent Systems*, *3*(1), 1–24. doi: 10.1145/2448116.2448118

Zeng, L., Proctor, R. W., & Salvendy, G. (2011). Can traditional divergent thinking tests be trusted in measuring and predicting real-world creativity? *Creativity Research Journal*, *23*(1), 24–37. doi: 10.1080/10400419.2011.545713

# Original publications

I    Oppenlaender, J., Milland, K., Visuri, A., Ipeirotis, P., & Hosio, S. (2020) Creativity on paid crowdsourcing platforms. *Proceedings of the 2020 CHI Conference on Human Factors in Computing Systems (CHI '20).* ACM: New York, NY, USA. 1–14. doi: 10.1145/3313831.3376677

II   Oppenlaender, J. & Hosio, S. (2019) Design recommendations for augmenting creative tasks with computational priming. *Proceedings of the 18th International Conference on Mobile and Ubiquitous Multimedia (MUM '19)*, ACM: New York, NY, USA. 35:1–35:13. doi: 10.1145/3365610.3365621

III  Oppenlaender, J., Kuosmanen, E., Lucero, A., & Hosio, S. (2021) Hardhats and Bungaloos: Comparing crowdsourced design feedback with peer design feedback in the classroom. *Proceedings of the 2021 CHI Conference on Human Factors in Computing Systems (CHI '21)*, ACM: New York, NY, USA. 1–14. doi: 10.1145/3411764.3445380

IV   Oppenlaender, J. & Hosio, S. (2019) Towards eliciting feedback for artworks on public displays. *Proceedings of the 2019 ACM Conference on Creativity & Cognition (C&C '19)*, ACM, New York, NY, USA. 562–569. doi: 10.1145/3325480.3326583

V    Oppenlaender, J., Kuosmanen, E., Goncalves, J., & Hosio, S. (2019) Search support for exploratory writing. Lamas, D., Loizides, F., Nacke, L., Petrie, H., Winckler, M., & Zaphiris, P. *Human-Computer Interaction – INTERACT 2019 (LNCS 11748)*, Springer: Cham, Switzerland. 314–336. doi: 10.1007/978-3-030-29387-1_18

VI   Oppenlaender, J., Tiropanis, T., & Hosio, S. (2020) CrowdUI: Supporting web design with the crowd. *Proceedings of the ACM on Human-Computer Interaction (PACMHCI '20).* Vol. 4, No. EICS, ACM: New York, NY, USA. 76:1–76:28. doi: 10.1145/3394978

778. Selkälä, Tuula (2021) Cellulose nanomaterials and their hybrid structures in the removal of aqueous micropollutants

779. Vilmi, Pauliina (2021) Component fabrication by printing methods for optics and electronics applications

780. Zhang, Kaitao (2021) Interfacial complexation of nanocellulose into functional filaments and their potential applications

781. Asgharimoghaddam, Hossein (2021) Resource management in large-scale wireless networks via random matrix methods

782. Kekkonen, Jere (2021) Performance of a 16×256 time-resolved CMOS single-photon avalanche diode line sensor in Raman spectroscopy applications

783. Borovkova, Mariia (2021) Polarization and terahertz imaging for functional characterization of biological tissues

784. de Castro Tomé, Mauricio (2021) Advanced electricity metering based on event-driven approaches

785. Järvenpää, Esko (2021) Yläpuolisen siltakaaren optimaalinen muoto

786. Lindholm, Maria (2021) Insights into undesired load factors at work now and tomorrow : findings from different professions and working conditions

787. Miettinen, Jyrki & Visuri, Ville-Valtteri & Fabritius, Timo (2021) Carbon-containing thermodynamic descriptions of the Fe–Cr–Cu–Mo–Ni–C system for modeling the solidification of steels

788. Tavakolian, Mohammad (2021) Efficient spatiotemporal representation learning for pain intensity estimation from facial expressions

789. Longi, Henna (2021) Regional innovation systems and company engagement in the Arctic context

790. Moltafet, Mohammad (2021) Information freshness in wireless networks

791. Nevanperä, Tuomas (2021) Catalytic oxidation of harmful chlorine- and sulphur-containing VOC emissions : a study of supported Au, Pt and Cu catalysts

792. Baharmast, Aram (2021) Wide dynamic range CMOS receiver techniques for a pulsed Time-of-Flight laser rangefinder

793. Haiko, Oskari (2021) Effect of microstructural characteristics and mechanical properties on the impact-abrasive and abrasive wear resistance of ultra-high strength steels

# A C T A   U N I V E R S I T A T I S   O U L U E N S I S

**UNIVERSITY OF OULU**